\documentclass[final,11p,times,twocolumn,floatfix,prd,authoryear,
nofootinbib,aps,eqsecnum,superscriptaddress,preprintnumbers]{revtex4-2}

\usepackage{bm}
\usepackage{xcolor}

\usepackage{newtxmath,newtxtext}
\usepackage{orcidlink}
\usepackage{graphicx}

\usepackage{hyperref}
\hypersetup{colorlinks=true,linkcolor=red,urlcolor=blue,	citecolor=red}

\newcommand{\bq}{\begin{equation}}
	\newcommand{\eq}{\end{equation}}
\newcommand{\bqn}{\begin{eqnarray}}
	\newcommand{\eqn}{\end{eqnarray}}

\def\Mpl{M_{\rm P}}

\begin{document}
\preprint{YITP-25-119, IPMU25-0041}

    \title{Dynamical dark energy parameterizations in VCDM}
    
	\author{Simran Arora \orcidlink{0000-0003-0326-8945}}
	\email{arora.simran@yukawa.kyoto-u.ac.jp}
	\affiliation{Center for Gravitational Physics and Quantum Information, Yukawa Institute for Theoretical Physics, Kyoto University, Kyoto, 606-8267, Japan}
    
	\author{Antonio De Felice \orcidlink{0000-0002-5556-4693}}
	\email{antonio.defelice@yukawa.kyoto-u.ac.jp}
	\affiliation{Center for Gravitational Physics and Quantum Information, Yukawa Institute for Theoretical Physics, Kyoto University, Kyoto, 606-8267, Japan}

\author{Shinji Mukohyama \orcidlink{0000-0002-9934-2785}}
	\email{shinji.mukohyama@yukawa.kyoto-u.ac.jp}
	\affiliation{Center for Gravitational Physics and Quantum Information, Yukawa Institute for Theoretical Physics, Kyoto University, Kyoto, 606-8267, Japan}
    \affiliation{Research Center for the Early Universe (RESCEU), Graduate School of Science, The University of Tokyo, Hongo 7-3-1, Bunkyo-ku, Tokyo 113-0033, Japan}
    \affiliation{Kavli Institute for the Physics and Mathematics of the Universe (WPI), The University of Tokyo Institutes for Advanced Study (UTIAS), The University of Tokyo, Kashiwa, Chiba 277-8583, Japan}

	\begin{abstract}
In the context of a theory of minimally modified gravity called VCDM, one can realize any cosmological behavior at the level of the homogeneous and isotropic background without introducing fatal instabilities for perturbations. Therefore, VCDM provides a theoretically-consistent and observationally-testable framework of dynamical dark energy parameterizations with or without phantom behaviors. In this paper, we propose the VCDM realizations of various phenomenological parameterizations present in the literature: the Chevallier–Polarski–Linder (CPL), Barboza–Alcaniz (BA), Jassal–Bagla–Padmanabhan (JBP), Exponential (EXP), and Logarithmic (LOG) models. Using the VCDM equations for cosmological perturbations, we test them against the recent cosmological datasets, Planck 2018 and DESI BAO DR2, and then discuss their implications. 
	\end{abstract}
	
	\maketitle
	\flushbottom
	
\section{Introduction}

Over the past two decades, a wealth of cosmological data has been gathered through numerous observational surveys, significantly enhancing the precision of cosmological parameter estimates. Our understanding of these observations has largely been shaped by the standard $\Lambda$-Cold Dark Matter ($\Lambda$CDM) model, which has served as a remarkably successful framework for interpreting the evolution of the Universe. While the $\Lambda$CDM model continues to offer an excellent fit to most current data, recent analyses of independent datasets have revealed a number of statistically significant tensions and anomalies that challenge its completeness.

A particularly prominent example of these tensions concerns the measurement of the Hubble constant, \( H_0 \), where a significant discrepancy exists between values inferred from the Cosmic Microwave Background (CMB) and those obtained through local distance ladder observations. Within the \(\Lambda\)CDM framework, Planck CMB data suggest a value of \( H_0 = 67.4 \pm 0.5 \,\textrm{km$\,$s$^{-1}$\,Mpc$^{-1}$} \) ~\cite{Planck:2018vyg}, while local, model-independent measurements from the Hubble Space Telescope (HST), based on 70 long-period Cepheid variables in the Large Magellanic Cloud, yield a considerably higher value of \( H_0 = 74.03 \pm 1.42 \,\textrm{km$\,$s$^{-1}$\,Mpc$^{-1}$} \)~\cite{Riess:2019cxk}. This difference amounts to a \( 4.4\sigma \) tension. Moreover, when incorporating time-delay cosmography from H0LiCOW gravitational lens systems alongside local distance ladder data, the tension with the Planck-based CMB estimate increases to \( 5.2\sigma \) \cite{H0LiCOW:2019pvv}, underscoring the growing challenge this poses to the standard cosmological model. This persistent and statistically significant discrepancy may point to unaccounted-for systematic effects or, more intriguingly, the need for new physics beyond the $\Lambda$CDM paradigm \cite{DiValentino:2021izs, Perivolaropoulos:2021jda, Abdalla:2022yfr, Kamionkowski:2022pkx, DiValentino:2020zio, Wu:2021jyk, Scolnic:2023mrv, Riess:2024vfa, Scolnic:2024hbh, Verde:2019ivm}.

One of the most straightforward extensions of the standard $\Lambda$CDM model involves introducing a dynamical dark energy (DDE) component, in which the dark energy equation of state (EoS), $w_{\text{de}}$, is either a constant different from $-1$ or varies with time. These so-called parametrized $w_{\text{de}}$ models provide a flexible framework for exploring potential departures from a cosmological constant \cite{Linder:2002et, Jassal:2004ej, Alam:2004jy, Pan:2017zoh, DeFelice:2012vd, DiValentino:2019dzu, Li:2019yem, Escamilla:2023oce, Giare:2024oil}. Recent results from the Year~1 survey of the Dark Energy Spectroscopic Instrument (DESI) offer compelling hints in this direction \cite{DESI:2024mwx}, suggesting, for the first time, a statistically significant deviation from $\Lambda$CDM. Specifically, within a DDE framework, the data show evidence of a time-varying $w_{\text{de}}(a)$ at greater than the $2\sigma$ confidence level. This finding has ignited active debate regarding its robustness and the broader implications for cosmology beyond $\Lambda$CDM \cite{Cortes:2024lgw, Park:2024vrw, Zheng:2024qzi, Nesseris:2025lke,Wu:2024faw}. 

The latest data release from DESI (DR2) \cite{DESI:2025zgx, DESI:2025fii, DESI:2025qqy} strengthens this trend, revealing a $3.1\sigma$ deviation from flat $\Lambda$CDM based on CMB and DESI data alone. Importantly, this preference for dynamical dark energy remains when supernova (SNe) datasets are incorporated, although the degree of significance varies depending on the specific sample used. For more reviews on DESI DR2, refer to \cite{Bhattacharjee:2025xeb,Plaza:2025gcv,Mishra:2025goj,Du:2025xes,Li:2025ops,Lee:2025kbn,Wang:2025znm,Li:2025dwz}. In this analysis, we focus exclusively on constraints derived from the CMB and the recent DESI DR2 measurements. We deliberately exclude Type Ia supernova data to minimize dependence on the cosmic distance ladder and its associated systematics, such as calibration uncertainties and potential redshift evolution in supernova luminosity. Additionally, CMB and BAO measurements are grounded in well-understood physical processes in the early Universe and large-scale structure, making them highly complementary and internally consistent within the $\Lambda$CDM framework and its extensions. 

In the present work, we restrict our analysis to a set of well-established parameterizations of the dark energy equation of state, \( w_{\text{de}} \), namely the Chevallier–Polarski–Linder (CPL), Barboza–Alcaniz (BA), Jassal–Bagla–Padmanabhan (JBP), Exponential (EXP), and Logarithmic (LOG) models, within the framework of a type-II minimally modified gravity theory known as VCDM \cite{DeFelice:2020eju}. An important advantage of VCDM is that it is free from ghost instabilities, thereby providing a theoretically consistent and stable foundation for exploring a wide range of dark energy dynamics. This ghost-free nature permits a broader and physically meaningful exploration of the parameter space. Moreover, our approach does not rely on prior assumptions regarding the dynamics or microphysical origin of the equation of state, enabling a purely observational investigation into whether the dark energy transition favors a quintessence-like or phantom-like behavior.

The structure of the paper is as follows. In Section~\ref{VCDM}, we provide a brief overview of the background of VCDM. Section~\ref{DDE} introduces the dynamical dark energy models considered in this study. Section~\ref{Data} describes the observational datasets employed and the methodology adopted for the analysis. In Section~\ref{results}, we present the resulting constraints on the model parameters. Finally, Section~\ref{conclusion} concludes the paper with a discussion of our findings and their implications.

\section{VCDM framework}
\label{VCDM}
In this study, we explore EoS models within the framework of minimally modified gravity, specifically the VCDM theory~\cite{DeFelice:2020eju}. In this approach, the cosmological constant $\Lambda$ in the standard $\Lambda$CDM model is generalized to a function $V(\phi)$, where $\phi$ is a non-dynamical auxiliary field. The `V' in VCDM represents the variable function $V(\phi)$ that is introduced in this framework. The VCDM minimal theory of gravity, being a minimal theory, does not introduce any additional local physical degrees of freedom, and it was built so as not to significantly alter the behavior of gravity and standard matter fields. In particular, Einstein's equivalence principle holds~\footnote{The total action of the system is the sum of the gravity action and the matter action, and the latter depends only on matter fields and a universal metric made of the lapse function, the shift vector and the spatial metric.}, gravitational waves still propagate with unity speed, and the effective gravitational constant for subhorizon cosmological modes is still equal to the Newton constant. These built-in features are meant to avoid issues with background stability, which often arise when attempting to model non-trivial cosmological background dynamics. Moreover, concerns about ghost instabilities are eliminated, as no extra gravitational modes are added beyond the standard tensor components. Consequently, this framework permits a wider range of phenomenological possibilities than conventional scalar-tensor theories for the background dynamics of a dark energy component. Several studies have been conducted to explore and analyze this framework in the contexts of cosmology \cite{DeFelice:2020cpt, Akarsu:2024qsi, Scherer:2025esj, Ganz:2022zgs, Ganz:2024ihb} and compact objects \cite{DeFelice:2020onz, DeFelice:2021xps, DeFelice:2022uxv, DeFelice:2022riv, Jalali:2023wqh}. 

The equations of motion within the VCDM theory can be expressed on a background that is both homogeneous and isotropic as:\footnote{Although the theory was initially formulated in the so-called unitary gauge \cite{DeFelice:2020eju}, it is possible to write down the VCDM theory in a covariant form by introducing a time-like St\"uckleberg field, $T$ \cite{DeFelice:2022uxv}. By means of this covariant form, one can derive all results known in the unitary gauge such as Eqs.\ \eqref{eq:Einst1}-\eqref{eq:cont_mat} or \eqref{VCDM1}-\eqref{eq:cont_mat-z}.}
\begin{align}
 V &= \frac13\,\phi^2 - \frac{\rho}{\Mpl^2}\,,\label{eq:Einst1}\\
 \frac{\dot\phi}{N}&=\frac32\frac{\rho+P}{\Mpl^2}\,.\label{eq:dotphi1}\\
\frac{\dot{\rho}_i}{N} &+ 3H(\rho_i + P_i)=0\,,\label{eq:cont_mat}
\end{align}
where a dot denotes differentiation with respect to time (later we chose conformal time $\tau$ by setting the lapse function $N(\tau)=a(\tau)$). The Hubble expansion rate is defined by $H = \frac{\dot{a}}{a N}$. The total energy density and pressure are expressed as $\rho = \sum_{i} \rho_{i}$, and $P = \sum_{i} P_{i}$, respectively, with the summation taken over all standard matter species, including dark matter. 

In order to implement a desired background dynamics for the dark energy component $\rho_{\rm de}\equiv\rho_{\rm de}(z)$, where, $z = a_{0}/a -1$ denotes the redshift (and the present-day scale factor is set to $a_{0} = 1$), we rewrite \eqref{eq:Einst1}-\eqref{eq:cont_mat} as 
\begin{align} 
\label{VCDM1}
3\Mpl^2 H^2 &= \rho+\rho_{\rm de}\,,\\
\rho_{\rm de} &= 3\Mpl^2 H^2+\Mpl^2\left(V-\frac13\,\phi^2\right)\,, \\
  (1+z)\frac{d\rho_{\rm de}}{dz} &= 3(\rho_{\rm de} + P_{\rm de})\,,\label{VCDM2}\\
  (1+z)\frac{d\rho_i}{dz} &= 3(\rho_i + P_i)\,.\label{eq:cont_mat-z}
\end{align}
At this point, once $\rho_{\rm de}(z)$ is given, the dark energy pressure is uniquely determined using Eq.\ \eqref{VCDM2}. Along the same lines, we uniquely determine $\frac{\dot{H}}{N}=-\frac{1+z}{2}\,\frac{d(H^2)}{dz}$.

The freedom in choosing the potential $V(\phi)$ allows one to reproduce a desired evolution of the Hubble parameter $H(z)$. Starting from the background equations and assuming $\rho + P > 0$ and $H>0$, one can determine the evolution of $\phi(z)$ and its inverse $z(\phi)$, thereby yielding a direct reconstruction of the potential $V(\phi)$. This approach developed in \cite{DeFelice:2020eju} (see also \cite{Ganz:2022zgs, Ganz:2024ihb} for its extension to the case without the assumption $H>0$ such as bouncing cosmology) provides an effective method for deriving the potential corresponding to a given cosmological background. In Appendix A, we perform the reconstruction by integrating the ODE that maps redshift $z$ to the field $\phi$, Eq.\ \eqref{eq:z_to_phi}, for concrete models studied in the following. Giving $\rho_{\rm de}(z)$, instead of defining $V(\phi)$ from the beginning, in an MCMC analysis, avoids the need to solve the ODE, Eq.\ \eqref{eq:dotphi1}, which sets how $\phi$ is mapped to time. From a theoretical perspective, the two possible ways of analyzing VCDM are equivalent, as far as the prior for the model parameters is the same.

In the Newtonian gauge,\footnote{In the unitary-gauge description of VCDM, there is no temporal gauge freedom and thus one cannot adopt the Newtonian gauge. Nonetheless, one can still take advantage of the use of variables that are defined in the same way as gauge-invariant variables in general relativity (GR). In addition to them, there are perturbations of Lagrange multipliers and one more variable stemming from the absence of the temporal gauge freedom, but these additional variables are consistently determined algebraically due to constraint equations present in the VCDM theory. In this way perturbation equations are written in terms of ``GR-gauge-invariant'' variables only. By ``the Newtonian gauge'' we mean the use of such ``GR-gauge-invariant'' variables that would reduce to the perturbation variables in the Newtonian gauge.} the equations of motion for cosmological perturbations remain identical to those in the $\Lambda$CDM model, with the sole exception of the momentum constraint, which takes the following exact form without any approximations~\footnote{Reflecting the fact that VCDM respects Einstein's equivalence principle, the perturbation equations depend on matter only through the total matter stress-energy tensor.}: 
\begin{equation}
    \dot{\Phi} + a H \Psi = \frac{3\left[k^2 - 3 a^2 (\dot{H}/a)\right] \sum_{i} (\varrho_{i} + p_{i}) \theta_{i} }{k^2 \left[ 2k^2 /a^2 + 9\sum_j (\varrho_{j} + p_{j}) \right]},
\end{equation}
where $\Phi$ and $\Psi$ are the two Bardeen potentials, and $\theta_i$ denotes the scalar perturbation of the fluid velocity for the $i$-th matter component. The indices $i$ and $j$ run over all standard matter species, including dark matter, but exclude the $\phi$-component. In the above equation, we chose the lapse to be $N=a$, so that a dot denotes differentiation with respect to the conformal time. Furthermore, we have defined $\varrho_i\equiv\rho_i/(3\Mpl^2)$ and $p_i\equiv P_i/(3\Mpl^2)$. In the limit $(\rho_{\rm de}+P_{\rm de})/(3\Mpl^2H^2)\to 0$, the momentum constraint reduces to the corresponding $\Lambda$CDM equation, ensuring the continuity of the $\Lambda$CDM limit. 

With the foundations of the VCDM theory laid out, we proceed to explore its implications by applying it to a range of $w(a)$ parameterizations, supported by observational data.

\section{Dynamical Dark Energy} \label{DDE}

One of the most effective approaches to modeling the behavior of the dark sector in the universe involves the parametrization of its respective components. This methodology has been extensively applied over the years to analyze dark energy fluid for a variety of purposes and observational studies. By considering a homogeneous and isotropic spacetime, characterized by the spatially flat Friedmann-Lemaître-Robertson-Walker (FLRW) metric, one can derive the Friedmann equations, which serve as solutions to the Einstein field equations. Consequently, the Hubble equation, which connects the total energy density of the universe with its expansion, can be formulated as follows:
\begin{equation}
\frac{H^2(z)}{H_0^2} = \Omega_{m0}(1 + z)^3 + \Omega_{r0}(1 + z)^4 + \left(1 - \Omega_{m0} - \Omega_{r0}\right) f_{\text{de}}(z)\,,
\label{H}
\end{equation}
where 
\begin{equation}
f_{\text{de}}(z) = \frac{\rho_{\text{de}}(z)}{\rho_{\text{de}0}} = \exp \left[ 3 \int_0^z \frac{1 + w(\tilde{z})}{1 + \tilde{z}} \, d\tilde{z} \right]\,.
\label{fde}
\end{equation}
The parameters \(\Omega_{m0}\) and \(\Omega_{r0}\) represent the current energy density of matter and radiation, respectively. These parameters evolve as a function of the scale factor \(a = \frac{1}{1+z}\) within the FLRW universe, under the assumption that \(a_0 = 1\). Furthermore, the Hubble parameter is defined as \(H(a) = \frac{\dot{a}}{a}\). The current value of the dark energy density is represented by $\rho_{\text{de}0}$, and selecting a specific form for $w(a)$ fulfills the criteria needed to ascertain the cosmological evolution dictated by Eq. \eqref{H}. There is considerable flexibility in selecting the function $w(a)$, which can be utilized to model the behaviors of various dark energy models.

We now introduce the specific forms of the models examined in this work. Our focus is on two-parameter models defined by the following free parameters: $w_0$, representing the present-day value of $w_{\text{de}}(a)$, and $
w_a = -\left.\frac{d w_{\text{de}}(a)}{d a}\right|_{a = a_0}$,
which captures the dynamical or non-dynamical behavior of $w_{\text{de}}(a)$. The standard $\Lambda$CDM model is recovered by setting $w_0 = -1$ and $w_a = 0$.

\begin{itemize}

\item Chevallier-Polarski-Linder (CPL): The CPL is the widely used parameterization \cite{Chevallier:2000qy, Linder:2002et}, where the dark energy equation of state is given by a Taylor expansion around $a = a_0 = 1$ up to the first-order term. In this framework, $w_0$ represents the current value of the equation of state, while $w_a$ characterizes its evolution with redshift. The form of $w(a)$ is:
\begin{equation}
w(a)=w_{0}+w_{a}(1-a)\,,
\end{equation}
for which
\begin{equation}
\rho_{\text{de}}(a)=\rho_{\text{de}0}\,a^{-3(w_{0}+w_{a})-3}{\mathrm{e}}^{3w_{a}(a-1)}\,.
\end{equation}

\item Barboza-Alcaniz (BA): This model has an EoS of the form:
\begin{equation}
w(a)=w_{0}+w_{a}\,\frac{1-a}{a^{2}+(1-a)^{2}}\,,
\end{equation}
such that
\begin{equation}
\rho_{\text{de}}(a)=\rho_{\text{de}0}\,\frac{\left(2a^{2}-2a+1\right)^{\frac{3w_{a}}{2}}}{a^{3(w_{0}+w_{a})+3}}\,.
\end{equation} 
The BA model improves upon the CPL parametrization by avoiding divergence as $a \to \infty$, ensuring a finite and well-defined behavior across the full expansion history of the universe \cite{Barboza:2008rh, Mehrabi:2018dru}. It permits phantom crossing and recovers $w_{\text{de}}(a) = w_0 + w_a$ in the limit $a \to 0$. Importantly, the model remains regular as $z \to -1$, which maps $a \in [0, \infty)$ to $z \in (-1, \infty)$. In addition to its stable high-redshift behavior, the BA model has been shown to reduce errors at low redshift more effectively than the CPL form.

\item Jassal-Bagla-Padmanabhan (JBP): This parameterization is a second-degree polynomial in $a$ \cite{Jassal:2004ej, Jassal:2005qc}, represented as:
\begin{equation}
w(a)=w_{0}+w_{a}\,a\,(1-a)\,,
\end{equation}
leading to
\begin{equation}  
\rho_{\text{de}}(a)=\rho_{\text{de}0}\,\frac{\exp\!\left[\frac{3w_{a}a(a-2)}{2}+\frac{3w_{a}}{2}\right]}{a^{3+3w_{0}}}\,.
\end{equation}
The model captures a dark energy component whose equation of state remains the same at both the present epoch and at high redshifts, while allowing for expected variations at low redshift. Since observational constraints are not very sensitive for $w(z)$ at $z \gg 1$, matching the present-day value in the distant past is not crucial. However, the model is well-suited for investigating differences in the dark energy dynamics at low redshift.

\item  Exponential (EXP): An alternative exponential parametrization may also be considered \cite{Pan:2019brc, Najafi:2024qzm}, defined by:
\begin{equation}
w(a)=w_{0}-w_{a}+w_{a} \mathrm{e}^{(1-a)}\,,
\end{equation}
with 
\begin{equation}
\rho_{\text{de}}(a)=\rho_{\text{de}0}\,\frac{e^{3w_{a}\,\mathrm{e}\,[\mathrm{Ei}_{1}(a)-\mathrm{Ei}_{1}(1)]}}{a^{3+3w_{0}-3w_{a}}}\,,
\end{equation}
where ${\rm Ei}_{1}(a)$ is the exponential integral function\footnote{This special function cannot be expressed in terms of elementary functions and is defined for $a>0$. For $a \leq 0$, it exhibits a logarithmic singularity. In our work, we utilized the GSL library, where the function \texttt{gsl\_sf\_expint\_E1} provides an implementation of this exponential integral.} ${\rm Ei}_{1}(a)={\rm Re}\!\left[\int_{1}^{\infty}{\rm d}t\,\exp(-a\,t)/t\right]$, and $0<a\leq1$. Expanding the exponential parametrization as a Taylor series reveals that it generalizes the CPL model by including terms beyond the linear approximation. However, as $a$ moves far away from $1$, the exponential form can introduce deviations, although small, from the linear CPL regime without increasing the dimensionality of the parameter space. 

\item  Logarithmic (LOG): The logarithmic term provides a smooth and monotonic evolution of $w(z)$, which can be useful in cosmological data fitting and avoids abrupt behavior \cite{Tripathi:2016slv}. This $w(a)$ form is defined as:
\begin{equation}
   w(a)=w_{0}-w_{a}\ln a\,, 
\end{equation}
which results in 
\begin{equation}
\rho_{\text{de}}(a)=\rho_{\text{de}0}\,a^{-3-3w_{0}}{\mathrm{e}}^{\frac{3w_{a}\ln^{2}a}{2}}.
\end{equation}
This equation of state was constructed empirically to fit certain quintessence models at redshifts $z \lesssim 4$ \cite{Efstathiou:1999tm}. However, supported by detailed stability analyses, we cautiously extend the validity of this parametrization to the high-redshift regime \cite{Yang:2021flj}, approaching $z \to \infty$.\footnote{In principle, for certain parameter combinations, the logarithmic term can grow significantly in absolute value and potentially lead to instabilities. However, given the current observational constraints and the slow growth of the logarithmic function, such behavior does not arise in our analysis. We find that the parametrization can be safely extended to high redshift, as the dark energy contribution remains subdominant compared to other components in the energy budget of the Universe.}

\end{itemize}

\section{Observational Data and Methodology}
\label{Data}

In this section, we briefly outline the cosmological datasets and the statistical methodology used to constrain the dynamical dark energy models considered in our VCDM framework. Specifically, we take into account the following:

\begin{itemize}
    \item CMB:  We make use of temperature and polarization anisotropy measurements of the CMB power spectra from the Planck satellite, in conjunction with their cross-spectra from the 2018 legacy data release of Planck. In particular, we apply the high-$\ell$ Plik likelihood for TT, TE, and EE, along with the low-l likelihood for TT only and the low-$\ell$ EE-only SimAll likelihood. We also include the reconstructed lensing potential obtained from the 3-point correlation function of the Planck data \cite{Planck:2018vyg, Planck:2019nip}.     
    \item DESI BAO: We consider the latest DESI BAO DR2 dataset, which includes observations of baryonic acoustic oscillations found in the clustering patterns of galaxies, quasars, and the Lyman-$\alpha$ forest at high redshifts, grouped into six distinct types of tracers \cite{DESI:2025fii, DESI:2025qqy, DESI:2025zgx}.  
\end{itemize}

We incorporate the dynamical dark energy model into our theoretical framework using the Boltzmann solver \texttt{CLASS} code~\cite{Blas:2011rf} and systematically examine the parameter space through Markov Chain Monte Carlo (MCMC) via the publicly available sampler \texttt{MontePython}~\cite{Brinckmann:2018cvx}. Parameter uncertainties and confidence contours are determined employing the \texttt{GetDist} package~\cite{Lewis:2019xzd}, which effectively processes the resulting Markov chains. We assess MCMC convergence using the Gelman–Rubin criterion, imposing $R-1 < 0.01$ to ensure reliable convergence. For our parameter inference, we use a broad uniform prior within the parameter space: \(\{ \Omega_b h^2,\, \Omega_c h^2,\, 100\,\theta_{s},\, \tau,\, n_s,\, \ln(10^{10}A_s),\, w_{0},\, w_{a}\}\). The first six parameters correspond to the standard cosmological parameters in the \(\Lambda\)CDM model, while the last two represent the parameters for dynamical dark energy. We present the constraints obtained by assuming five parametrizations for the DE EoS in Table~\ref{tab:model_params}. The VCDM framework, as a minimal and theoretically consistent extension of $\Lambda$CDM, offers a broader parameter space for exploring cosmic dynamics. Its freedom from ghosts and instabilities ensures stability at both background and perturbative levels, allowing for a wider range of viable cosmological behaviors while naturally including $\Lambda$CDM as a limiting case.

\begin{figure}[]
\centering
\includegraphics[scale=0.45]{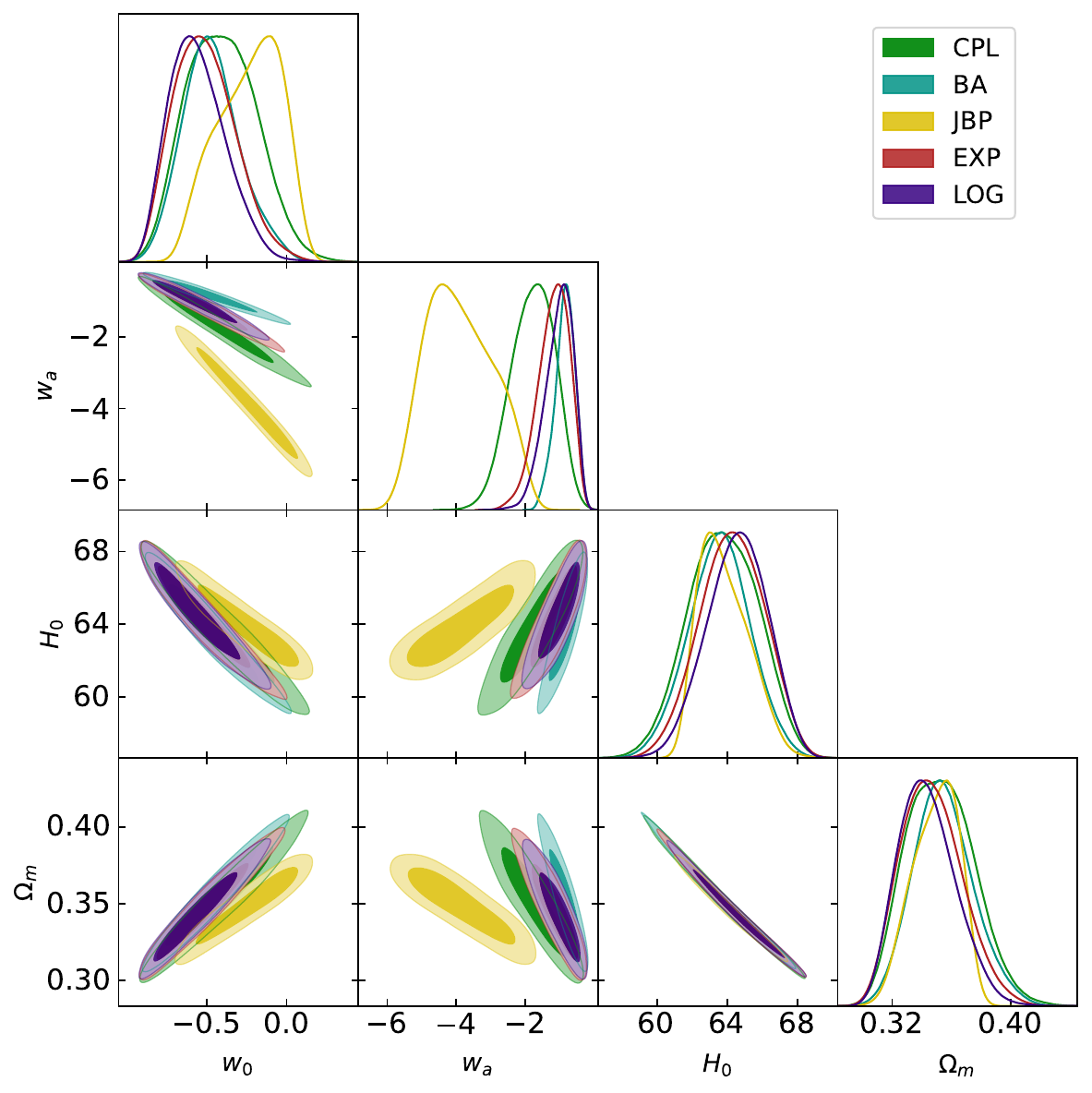}
\caption{One-dimensional marginalized posterior distributions and two-dimensional joint contours of the model parameters considering $CMB + DESI$ in VCDM.}
\label{fig:corner_plots}
\end{figure}

To quantify the preference of the observational datasets for the dark energy models presented in section \ref{DDE}, relative to $\Lambda$CDM, we employ two statistical measures: the change in the minimum (best-fitting) chi-squared value, $\Delta \chi^2_{\rm min} = \chi^2_{\mathrm{min}}(\Lambda\mathrm{CDM})-\chi^2_{\mathrm{min}}(\mathrm{model})$, and model selection criteria, namely the Akaike Information Criterion (AIC) defined as $\mathrm{AIC} \equiv -2 \ln \mathcal{L}_{\max} + 2k$, where $\mathcal{L}_{\max}$ is the maximum likelihood of the model, $k$ is the number of free parameters. Consequently, $\Delta AIC$ is defined as $\Delta AIC =AIC (\Lambda\mathrm{CDM})- AIC (\mathrm{model})$. A lower $AIC$ value indicates a more favorable model, as it provides a balance between the goodness of fit and the complexity of the model.
To identify the best-fit parameters and corresponding $\chi^2_{\rm min}$, we utilized the \textit{minimize.py} routine provided in the MontePython package. Specifically, we followed the implementation available here,\footnote{\url{https://github.com/GuillermoFrancoAbellan/MontePython/blob/main/run_minimizer.sh}} which employs optimization algorithms.

As the results will show, the value of $H_0$ is not consistent with the SH0ES data. If the SH0ES data are reliable, then the models discussed here cannot simultaneously provide a good fit to the PLANCK, DESI, and SH0ES observations, and alternative dark energy models must be considered. However, the physics underlying PLANCK and BAO-bump measurements is likely better understood than the complex, non-linear stellar evolution involved in modeling the astrophysical sources behind supernova data. For this reason, we adopt a conservative approach in this paper and do not include any supernova data sets. Instead, we rely entirely on cosmological observations related to the background evolution and perturbation theory. Upcoming data, including gravitational wave measurements, are expected to shed further light on the $H_0$ tension in the coming years.

\begin{figure*}

\includegraphics[scale=0.35]{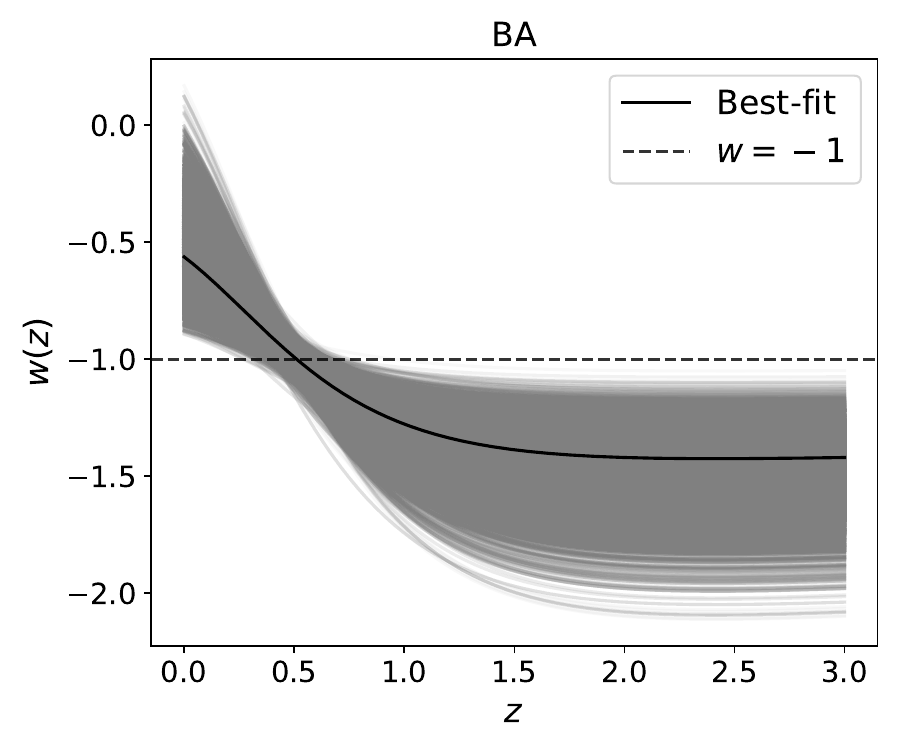} 
\hspace{0.01in}
\includegraphics[scale=0.35]{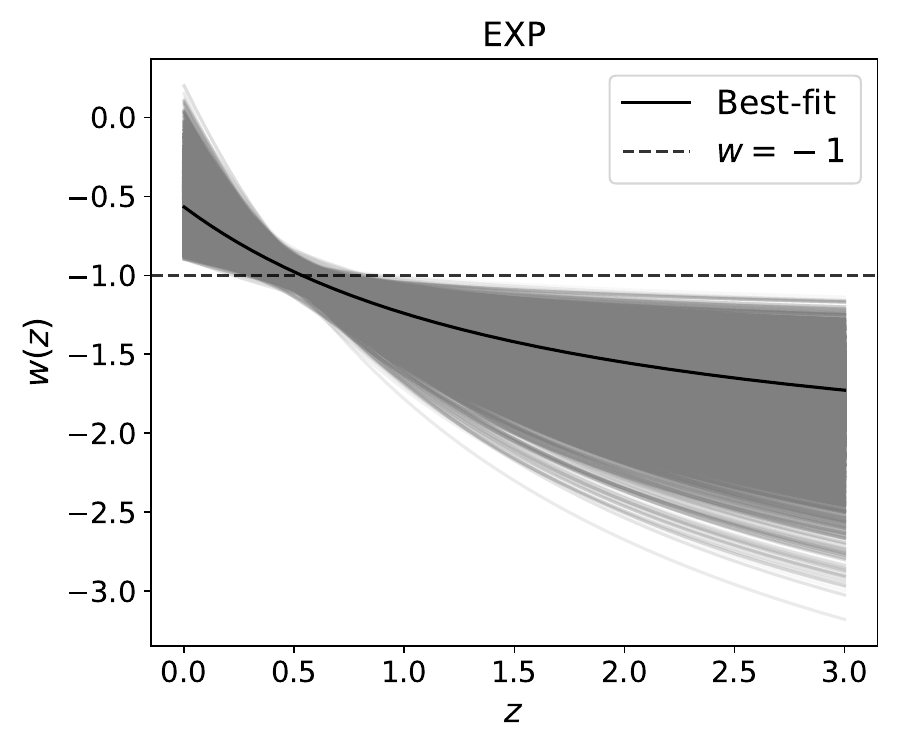}
\hspace{0.01in}
\includegraphics[scale=0.35]{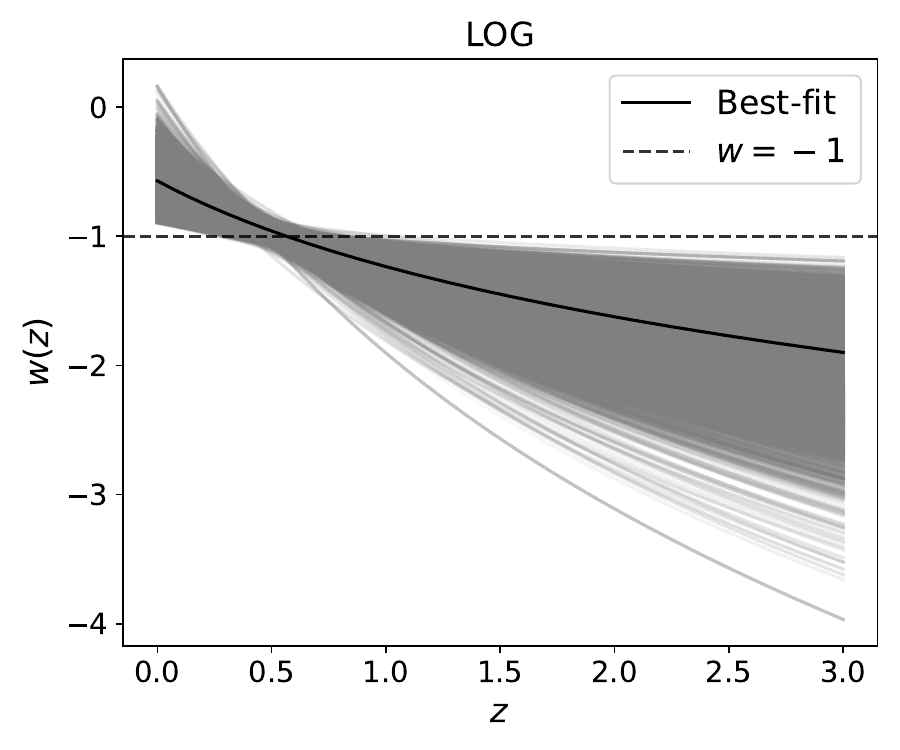}
\hspace{0.01in}
\includegraphics[scale=0.35]{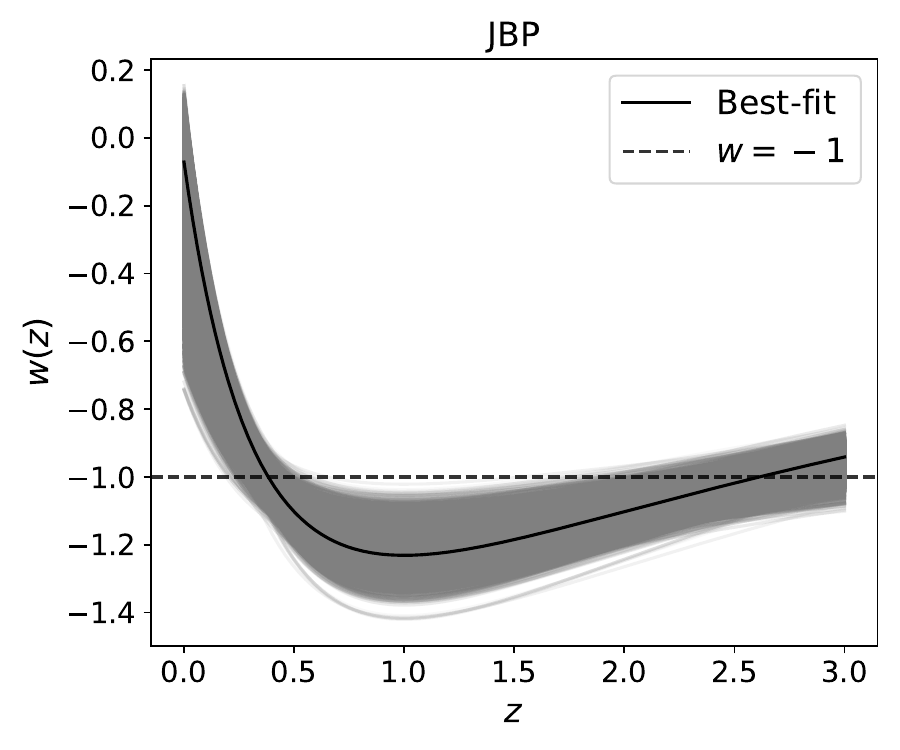}
\hspace{0.01in}
\includegraphics[scale=0.35]{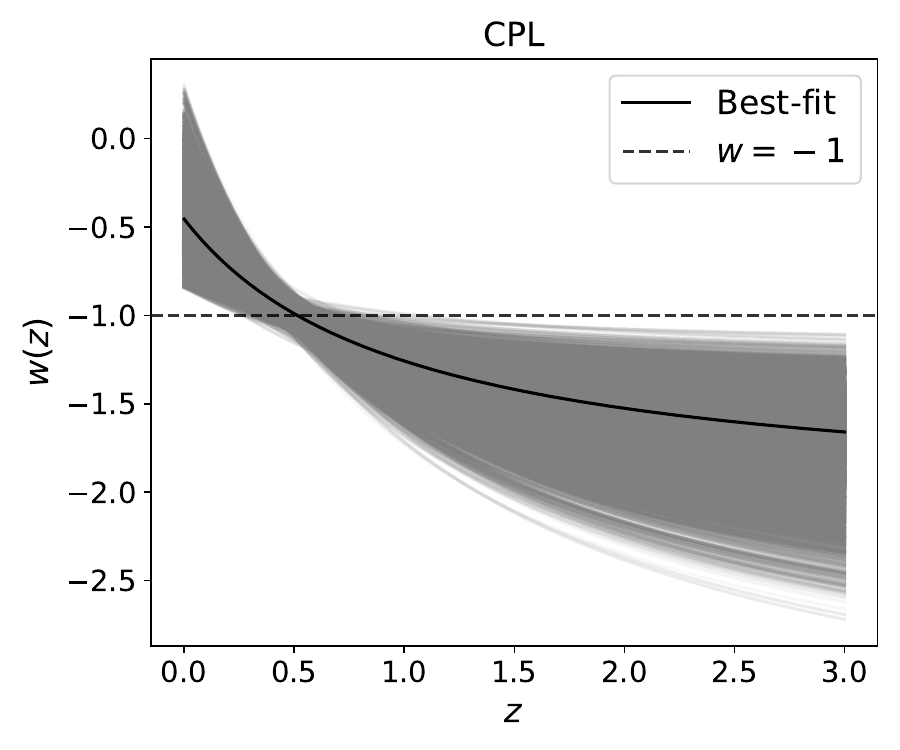}
\caption{The evolution of the equation of state parameter with respect to redshift $z$ for all the five parametrizations.}
\label{fig:w}

\end{figure*}

\begin{table*}[]
\centering
\begin{tabular}{lcccccc}
\hline
\multicolumn{6}{c}{\textbf{68\% confidence level (CL) constraints for the parameters ({\color{blue}Best-fit value})}} \\
\hline
Parameters &  $CPL$ & $BA$ & $JBP$ & $EXP$  & $LOG$ & $\Lambda CDM$\\
 \hline \hline
 $10^2\Omega_{b}h^2$  & $2.239 \pm 0.014 $ & $2.239 \pm 0.013$ & $2.245 \pm 0.013$ & $2.239 \pm 0.014 $ & $2.238 \pm 0.014 $  & $2.254 \pm 0.013 $ \\ [0.15em]
   & {\color{blue} $2.242$} & {\color{blue}$2.242$} & {\color{blue}$2.246$} & {\color{blue}$2.242$} & {\color{blue}$2.236$} & {\color{blue} $2.255$}   \\ [0.15em]
  $\Omega_{c}h^2$  & $0.11974 \pm 0.00094 $ & $0.1196
  \pm 0.00093 $ & $0.11896 \pm 0.00087$ & $0.11978 \pm 0.00094$ & $0.11982 \pm 0.00094$ & $0.11771 \pm 0.00066$   \\ [0.15em]
     & {\color{blue}$0.11967$} & {\color{blue}$0.11955$} & {\color{blue}$0.11896$} & {\color{blue}$0.11955$} & {\color{blue}$0.11992$} & {\color{blue}$0.11775$}   \\ [0.15em]
    $100\theta_{s}$  & $1.04189 \pm 0.00028 $ & $1.04191 \pm 0.00028 $ & $1.04200 \pm 0.00028 $ & $1.04191 \pm 0.00029 $ & $1.04190 \pm 0.00029 $  & $1.04213 \pm 0.00027 $  \\ [0.15em]
      & {\color{blue}$1.04191$} & {\color{blue}$1.04196$} & {\color{blue}$1.041967$} & {\color{blue}$1.04190$} & {\color{blue}$1.04194$} & {\color{blue}$1.04209$}  \\ [0.15em]
    $ln(10^{10} A_s)$ & $3.041 \pm 0.014 $ & $3.041 \pm 0.014 $ & $3.048 \pm 0.014 $ & $3.042 \pm 0.014$ & $3.040 \pm 0.014$  & $3.054^{+0.015}_{-0.016}$ \\ [0.15em]
    & {\color{blue}$3.042$} & {$\color{blue}3.041$} & {\color{blue}$3.045$} & {\color{blue}$3.044$} & {\color{blue}$3.036$} & {\color{blue}$3.052$} \\ [0.15em]
$w_{0}$  & $-0.40^{+0.19}_{-0.26} $ & $-0.51 \pm {0.17} $ & $-0.23^{+0.25}_{-0.16} $ & $-0.51^{+0.17}_{-0.22} $ & $-0.56^{+0.14}_{-0.21}$ & - \\  [0.15em]
  & {\color{blue}$-0.45$} & {\color{blue}$-0.563$} & {\color{blue}$-0.0714$} & {\color{blue}$-0.567$} & {\color{blue}$-0.571$} & -  \\  [0.15em]
$w_{a}$   & $-1.78^{+0.77}_{-0.54} $ & $-0.81^{+0.30}_{-0.24} $ & $-3.89^{+0.85}_{-1.2} $ &  $-1.19^{+0.55}_{-0.57} $ & $-1.03^{+0.40}_{-0.30}$  & - \\ [0.15em]
& {\color{blue}$-1.605$} & {\color{blue}$-0.714$} & {\color{blue}$-4.639$} & {\color{blue}$-1.038$} & {\color{blue}$-0.958$} & - \\ [0.15em]
$H_{0}$  & $63.6 \pm 1.9 $ & $63.9 \pm 1.7$ & $63.8 ^{+1.1}_{-1.5}$ & $64.3 \pm 1.8 $ & $64.6^{+1.9}_{-1.6}$ & $68.44 \pm 0.30$ \\ [0.15em]
& {\color{blue}$63.9$} & {\color{blue}$64.4$} & {\color{blue}$62.6$} & {\color{blue}$64.6$} & {\color{blue}$64.3$} & {\color{blue}$68.42$} \\ [0.15em]
$\Omega_{m0}$  & $0.354^{+0.020}_{-0.026} $ & $0.35^{+0.018}_{-0.022}$ & $0.349^{+0.015}_{-0.014} $ & $0.347^{+0.018}_{-0.023} $ & $0.344^{+0.016}_{-0.022} $ & $0.301 \pm 0.0038$   \\ [0.15em]
 & {\color{blue}$0.349$} & {\color{blue}$0.344$} & {\color{blue}$0.362$} & {\color{blue}$0.342$} & {\color{blue}$0.345$}  & {\color{blue}$0.301$} \\ [0.15em]
$S_{8}$  & $0.847^{+0.014}_{-0.015} $ & $0.844 \pm {0.013} $ & $0.837 \pm {0.010}$ & $0.845 \pm {0.014}$ & $0.844 \pm {0.013}$ & $0.8099 \pm {0.0084}$ \\  [0.15em]
 & {\color{blue}$0.845$} & {\color{blue}$0.841$} & {\color{blue}$0.840$} & {\color{blue}$0.842$} & {\color{blue}$0.843$} & {\color{blue}$0.8099$} \\  [0.15em]
$n_{s}$  & $0.9659 \pm 0.0037 $ & $0.9661 \pm 0.0036 $ & $0.9679 \pm 0.0036$ & $0.9661 \pm 0.0036 $ & $0.9657 \pm 0.0037 $  & $0.9710 \pm 0.0034 $\\ [0.15em]
 & {\color{blue}$0.9664$} & {\color{blue}$0.9679$} & {\color{blue}$0.9686$} & {\color{blue}$0.9677$} &  {\color{blue}$0.9652$} & {\color{blue}$0.9720$}  \\ [0.15em]
$\tau_{reio}$  & $0.0530 \pm 0.0072 $ & $0.0530 \pm 0.0071 $ & $0.0568 \pm 0.0072$ & $0.0532 \pm 0.0072$ & $0.0526 \pm 0.0073$ & $0.0613^{+0.0087}_{-0.0077}$  \\
  & {\color{blue}$0.0537$} & {\color{blue}$0.0532$} & {\color{blue}$0.0559$} & {\color{blue}$0.0543$} & {\color{blue}$0.0514$} & {\color{blue}$0.060$}\\ \hline \hline
$\chi^{2}_{\rm min,planck}$ & $2772.97$ & $2773.02$ & $2774.65$ & $2772.89$ & $2774.27$ & $2777.65$ \\ [0.3em] 
$\chi^{2}_{\rm min,DESI}$ & $7.1$ & $6.87$ & $8.25$ & $7.25$ & $7.66$ & $11.87$ \\ [0.3em] 
$\Delta \chi^{2}_{\rm min}$ & $9.44$ & $9.62$ & $6.62$ & $9.38$ & $7.58$ & $0$ \\ [0.3em]
$\Delta AIC$ & $5.44$ & $5.62$ & $2.62$ & $5.38$ & $3.58$  & $0$\\  \hline
\hline
\end{tabular}
\caption{Marginalized constraints and mean values with 68\% CL on the free and some derived parameters of the $w(z)$ parameterizations in the VCDM framework from the combinations of the DESI DR2 and Planck 2018 CMB datasets. We present $\Delta \chi^2_{\rm min}$ and $\Delta AIC$ values for the best fit parameters.
}
\label{tab:model_params}
\end{table*}

\section{Results and Discussions}
\label{results}

In this section, we present and discuss the results of our analysis. Fig.~\ref{fig:corner_plots} displays the marginalized constraints with 68\% and 95\% confidence regions for the VCDM framework with various dark energy parametrizations: CPL, BA, JBP, EXP, and LOG for CMB + DESI DR2 data combination. The parameters constrained include the dark energy equation of state parameters $w_0$ and $w_a$, the Hubble constant $H_0$, and the matter density parameter $\Omega_m$. Constraints on additional cosmological parameters are summarized in Table~\ref{tab:model_params}.

Compared to the standard $\Lambda$CDM model, defined by $w_0 = -1$ and $w_a = 0$, the VCDM parametrizations allow for a dynamical dark energy sector. This added flexibility results in broader and more structured posterior distributions across the parameter space. While all models produce comparable constraints in the $H_0$--$\Omega_m$ plane, with largely overlapping credible regions, significant differences emerge in the $w_0$--$w_a$ plane. In particular, the JBP model exhibits broader and more tilted contours, indicative of greater freedom but weaker constraining power relative to the other parameterizations.

The combined data favor the region $w_0 > -1$ and $w_a < 0$, deviating from the cosmological constant scenario. This implies that the dark energy equation of state was phantom-like in the distant past and has evolved to $w(z) > -1$ at present, as illustrated in Fig.~\ref{fig:w}. A similar trend has been reported in the recent DESI analysis; however, our results reaffirm this behavior within the theoretically consistent framework of VCDM.

Fig.~\ref{fig:w} presents the redshift evolution of the dark energy equation of state \( w(z) \) for several parametrizations within the VCDM framework, based on 10,000 samples extracted from the MCMC chains using combined CMB and DESI DR2 data. The solid black curves represent the best-fit trajectories corresponding to the minimum chi-square values for each model. Among the parametrizations considered, the LOG, EXP, and CPL models exhibit significant dynamical evolution, with \( w(z) \) deviating markedly from the cosmological constant value \( w = -1 \) at higher redshifts. These models accommodate greater flexibility in the equation of state, as evidenced by the broader spread of \( w(z) \) at early times. In contrast, the JBP and BA models display more moderate evolution, with \( w(z) \) remaining closer to \(-1\) over the full redshift range. The JBP model, in particular, features a distinctive turnaround behavior: \( w(z) \) undergoes a sharp decline at low redshift followed by a gradual rise, although its overall variation remains comparatively restrained.

The combined DESI and CMB datasets impose stringent constraints on the dark energy equation of state $w(z)$, particularly at low redshifts ($z \leq 1$), where the influence of dark energy on the expansion history is most pronounced. This is evident from the narrower uncertainty bands across all parametrizations, indicating robust constraints in this regime. At higher redshifts, however, $w(z)$ becomes progressively less constrained, reflecting both the diminished impact of dark energy and the reduced precision of current observations. Among the parametrizations, the CPL, EXP, and BA models exhibit a transition from $w > -1$ to $w < -1$ near $z \sim 0.5$, consistent with the standard redshift associated with the onset of cosmic acceleration. In contrast, the JBP model indicates a delayed transition, around $z \sim 0.3$–$0.4$, while the LOG model displays a more rapid evolution, with marked deviations from $w = -1$ appearing beyond $z \sim 0.5$. These variations underscore the sensitivity of reconstructed dark energy dynamics to the choice of parametrization.

\begin{figure}[]
\centering
\includegraphics[scale=0.4]{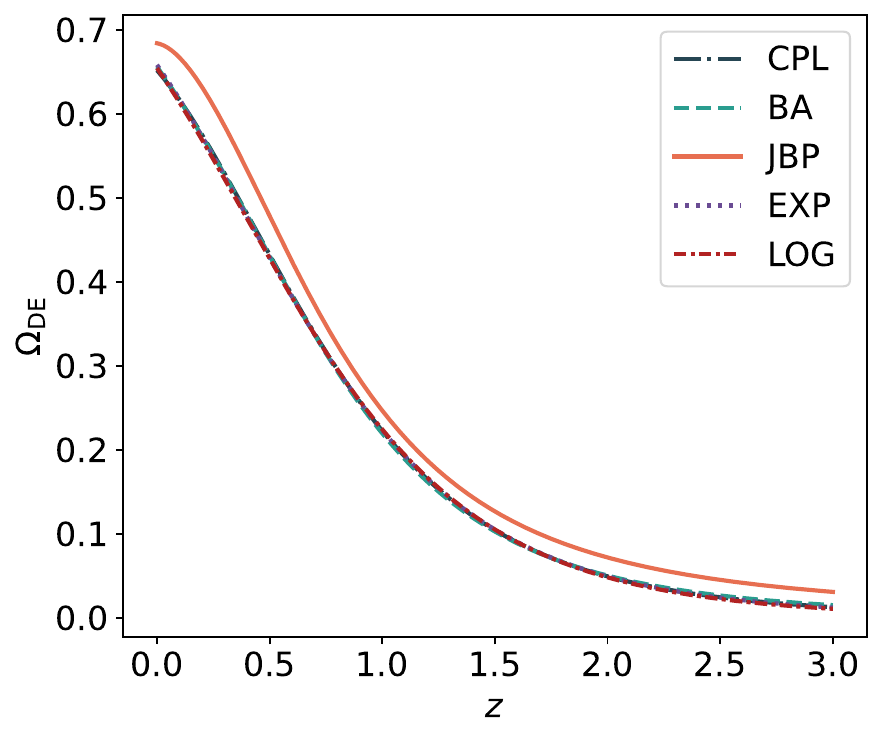} 
\caption{The evolution of the dark energy density parameter with respect to redshift $z$ for all five parameterizations using best-fit values.}
\label{fig:Omega}

\end{figure}

Fig.~\ref{fig:Omega} shows the redshift evolution of the dark energy density parameter $\Omega_{\rm DE}(z)$ for five different $w(z)$ parametrizations within the VCDM framework, constrained using the combined DESI and CMB data. All models exhibit the expected monotonic decrease in $\Omega_{\rm DE}(z)$ with increasing redshift, consistent with dark energy becoming subdominant in the early Universe. At low redshifts ($z \lesssim 1$), the curves remain tightly clustered, indicating that the observational data strongly constrain the evolution of dark energy in this regime. At higher redshifts ($z \gtrsim 1.5$), slight differences appear, with the JBP model showing marginally higher values of $\Omega_{\rm DE}(z)$, a consequence of its transition behavior in $w(z)$. Despite these differences, all models converge to $\Omega_{\rm DE} \to 0$ by $z \sim 3$, consistent with the matter-dominated era of the early Universe.

In comparing the dynamical dark energy models with the $\Lambda$CDM model, we note that the former introduce two additional parameters: $w_0$ and $w_a$. As a result, $\Lambda$CDM can be regarded as a nested model within the broader class of dynamical models, such as those considered in the VCDM framework. Consequently, the difference in the AIC simplifies to $\Delta \mathrm{AIC} = \Delta \chi^2_{\mathrm{min}} - 4$ (as, $\Delta k = -2)$. For the dynamical models to be statistically favored under this criterion, it is necessary that $\Delta \chi^2_{\mathrm{min}} > 4$. As evident from Table~\ref{tab:model_params}, this condition is clearly satisfied.
In this context, the CPL, BA, and EXP models yield, which provides strong statistical support for these models over the standard $\Lambda$CDM scenario. This indicates that the additional degrees of freedom introduced in the dark energy sector are not only statistically justified but also lead to a significantly improved fit to the observational data. The Logarithmic model exhibits a moderate improvement, with $\Delta AIC \simeq 3.6$, which may be interpreted as weak-to-moderate evidence in its favor. In contrast, the JBP model, with $\Delta AIC \simeq 2.6$, offers only a marginal improvement over $\Lambda$CDM. 

\begin{figure}[]
\centering
\includegraphics[scale=0.4]{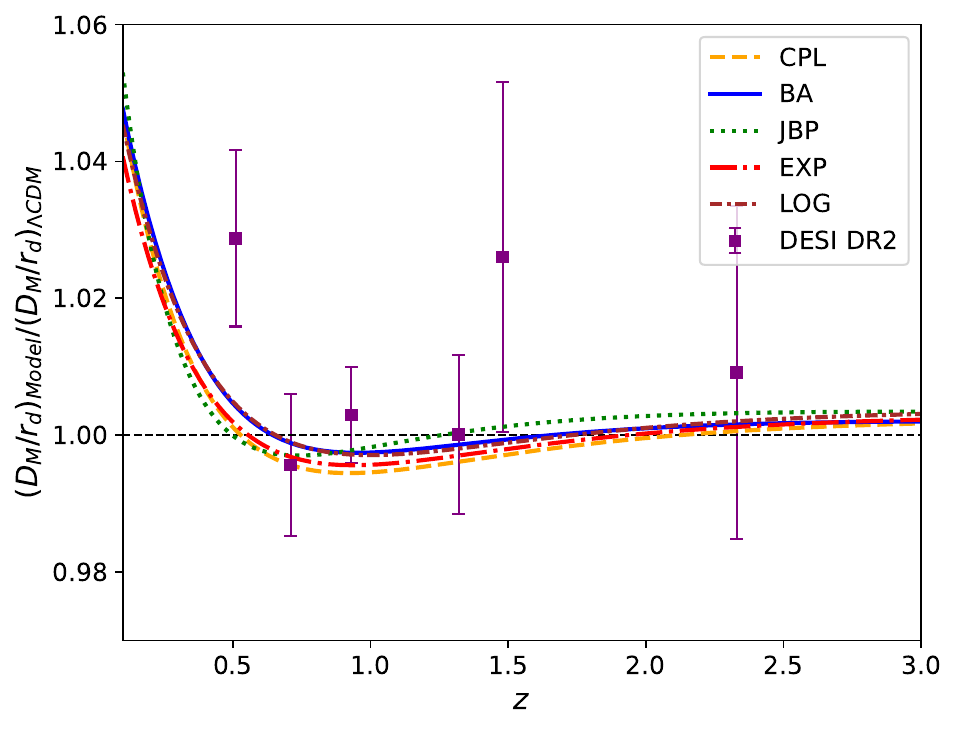} 
\hspace{0.01in}
\includegraphics[scale=0.4]{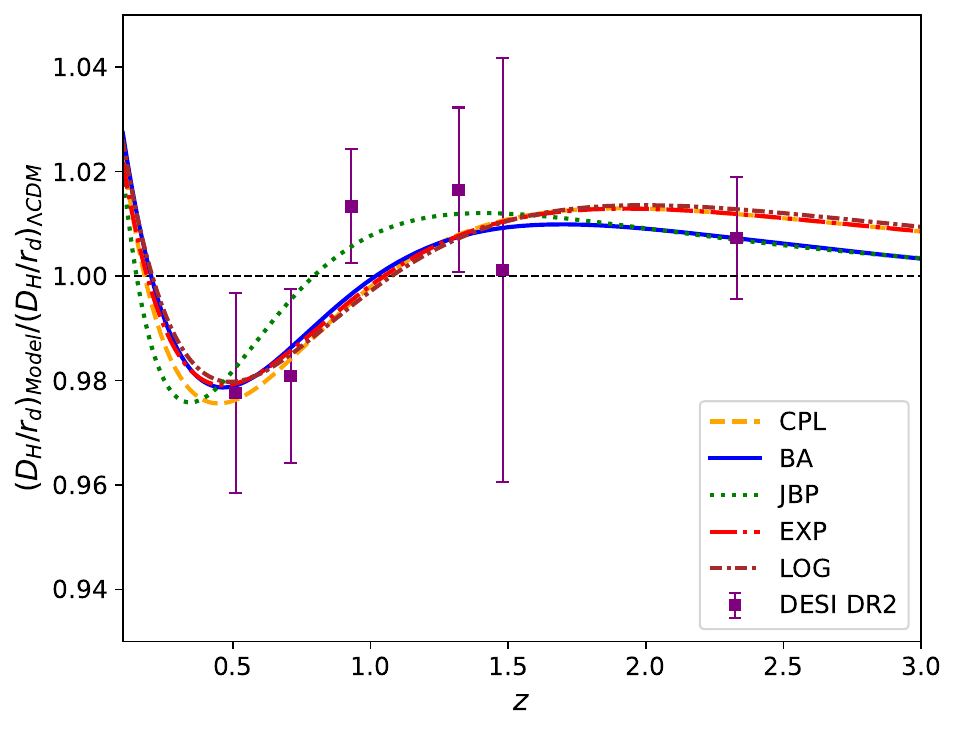}
\hspace{0.01in}
\caption{The BAO distance ratios \( (D_M/r_d)/(D_M/r_d)_{\Lambda\mathrm{CDM}}\) for the CPL, BA, JBP, EXP, and LOG models using DESI DR2 + Planck data. These distances are normalized by the best-fit \(\Lambda\mathrm{CDM}\) model derived from the same dataset. DESI DR2 measurements are shown as purple points with error bars.}
\label{fig:dm/dh}

\end{figure}

To further interpret our findings, we reconstruct the normalized BAO distance ratios, \( (D_M/r_d)_{\text{model}} / (D_M/r_d)_{\Lambda\text{CDM}} \) and \( (D_H/r_d)_{\text{model}} / (D_H/r_d)_{\Lambda\text{CDM}} \), for a range of dynamical dark energy parametrizations—CPL, BA, JBP, EXP, and LOG. The curves correspond to best-fit values obtained within the VCDM framework, using constraints from the combined Planck 2018 and DESI DR2  datasets in Fig.~\ref{fig:dm/dh}. Both sets of reconstructed distances exhibit a consistent, non-monotonic deviation from the standard \(\Lambda\)CDM prediction. For the transverse comoving distance, \( D_M/r_d \), all models show a mild enhancement at low redshifts (\( z \lesssim 0.3 \)), a suppression at intermediate redshifts (\( z \sim 0.5 - 1.2 \)), and a return to slightly elevated values at higher redshifts (\( z \gtrsim 2 \)). For the Hubble distance, \( D_H/r_d \), a similar trend is observed, with a suppression at low redshifts and a gradual recovery to \(\Lambda\)CDM-like or slightly higher values beyond \( z \sim 1.5 \). Among the parametrizations, the BA and JBP models tend to show more pronounced deviations near \( z \sim 1 \). Despite these variations, all predictions remain within the $1\sigma$ bounds of the DESI DR2 measurements, thereby hinting at a preference for dynamical dark energy.

Since all the analyses above exhibit a consistent overall trend within the VCDM framework, it is important to emphasize that the VCDM model not only corroborates the DESI findings and improves the overall goodness of fit but also provides a robust theoretical foundation for the various dark energy parameterizations considered. Although our model offers a better fit to the data compared to the standard $\Lambda$CDM scenario, it does not resolve the $H_0$ tension nor significantly alleviate the $S_8$ discrepancy \footnote{Although the present work does not primarily focus on the Hubble tension, readers interested in a broader perspective on $H_{0}$ and its recent work may refer to the following references \cite{Wang:2025bkk,Poulin:2025nfb,Ling:2025lmw,Lee:2022cyh}.}. Nevertheless, the improved fit and consistency with late-time observational data may suggest that the VCDM framework captures key features of the dark sector evolution, potentially pointing toward new physics responsible for the late-time cosmic acceleration.

\section{Conclusions}

\label{conclusion}

The cosmological constant has long served as a central pillar of the standard $\Lambda$CDM model. However, recent developments have prompted growing interest in dark energy models with a time-dependent equation of state. In particular, the latest results from DESI Data Release 2 motivate a closer examination of such evolving scenarios. Emerging tensions and potential anomalies in the dark energy sector have become subjects of active discussion, underscoring the need for further scrutiny using high-precision cosmological data.

In this context, recent advances in BAO measurements from DESI, when combined with Planck 2018 observations of the CMB, provide tentative evidence for a dynamical dark energy component. The inferred behavior resembles quintessence at present but allows for a crossing of the phantom divide $w = -1$ at higher redshifts. This is particularly notable, as such phantom crossing is generally precluded in conventional quintessence models, suggesting that alternative frameworks, possibly beyond standard scalar field dynamics, may be required to account for the data. \footnote{If we choose a canonical scalar field with a wrong sign kinetic term to realize the phantom behavior, then the perturbations are completely unstable everywhere and every time. On the other hand, if we choose a canonical scalar field with a correct sign kinetic term, then the phantom behavior is impossible.}

In the current era of precision cosmology, growing inconsistencies among various datasets and between observations and the predictions of GR have sparked renewed interest in the potential need for modifications to the gravitational sector. Given that no observational evidence has thus far necessitated the introduction of additional gravitational degrees of freedom at small scales, it is well motivated to focus on models that preserve the two tensorial polarizations of gravitational waves. This class of theories, known as minimally modified gravity (MMG), offers a compelling framework for such investigations.
In this work, we explore such models through the lens of the VCDM scenario, with redshift-dependent parameterizations of the dark energy equation of state, $w(z)$. By confronting these models with observational data from DESI DR2 and the Planck 2018 CMB measurements, we derive new constraints on $w(z)$. Our analysis indicates that the tightest constraints emerge at low redshifts, where $w(z)$ mildly favors deviations with $w > -1$, while at higher redshifts, the results are consistent with a transition into the phantom regime ($w(z) < -1$), suggesting potential dynamical features in the dark energy sector.

Among the parametrizations studied, the CPL, EXP, and BA models exhibit a transition from $w > -1$ to $w < -1$ near $z \sim 0.5$, aligning with the expected onset of cosmic acceleration. The JBP model shows a comparatively delayed transition around $z \sim 0.3-0.4$, while the LOG model reveals a more rapid evolution, with notable deviations from $w = -1$ beyond $z \sim 0.5$. 
Additionally, we conducted a comparative analysis against the standard $\Lambda$CDM model using both $\chi^2$ and AIC. The CPL, BA, and EXP models demonstrate strong statistical preference, indicating that the inclusion of additional degrees of freedom in the dark energy sector is both statistically warranted and results in a significantly improved fit to the observational data. The LOG model shows a moderate improvement, providing weak-to-moderate evidence in its favor, whereas the JBP model offers only marginal support. Embedding these parameterizations within a fully predictive model, while allowing phantom crossing without introducing theoretical inconsistencies, is one of the main results of our paper.

Our results open promising avenues for probing the nature of dark energy beyond the conventional cosmological constant framework. Importantly, the inferred redshift range for the dark energy transition lies well within the reach of ongoing observational programs such as DESI. Looking ahead, forthcoming CMB experiments are expected to significantly tighten constraints on early-universe parameters, helping to break key degeneracies with late-time cosmological observables. Together, these developments set the stage for a transformative decade in cosmology, one that may ultimately challenge and reshape our current theoretical paradigm.

\section*{Acknowledgments}
\noindent SA acknowledges the Japan Society for the Promotion of Science (JSPS) for providing a postdoctoral fellowship during 2024-2026 (JSPS ID No.: P24318). This work of SA is supported by the JSPS KAKENHI grant (Number: 24KF0229). The work of SM was supported in part by JSPS KAKENHI Grant No.\ JP24K07017 and World Premier International Research Center Initiative (WPI), MEXT, Japan. We thank the editor and the referee for their constructive and insightful comments, which have significantly improved our manuscript. 

\appendix*
\section{Reconstruction of the VCDM potential}
\begin{figure*}

\includegraphics[scale=0.35]{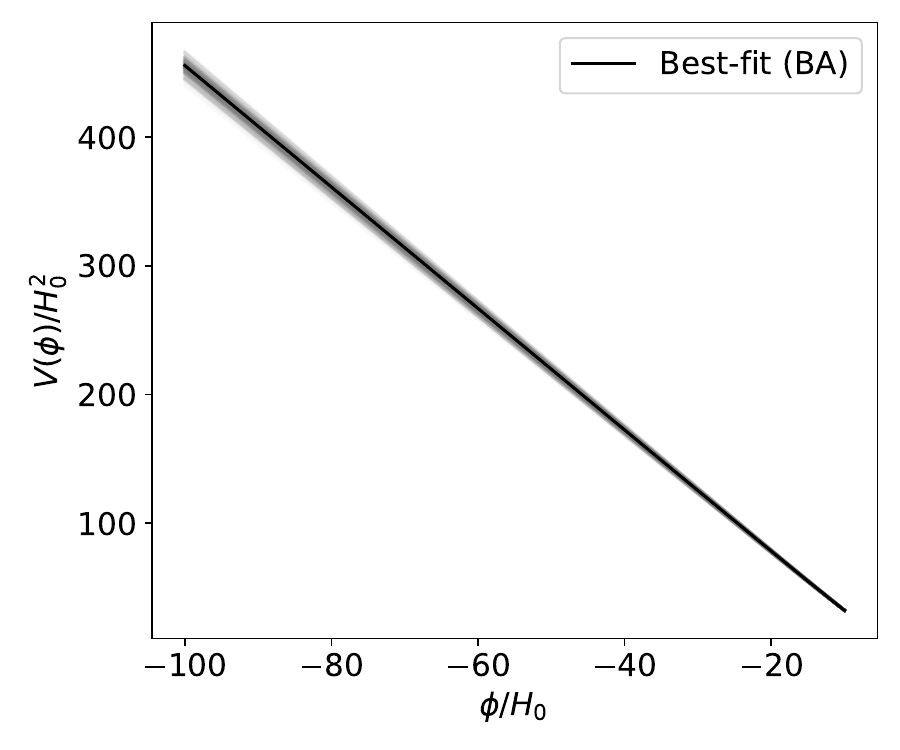} 
\hspace{0.01in}
\includegraphics[scale=0.35]{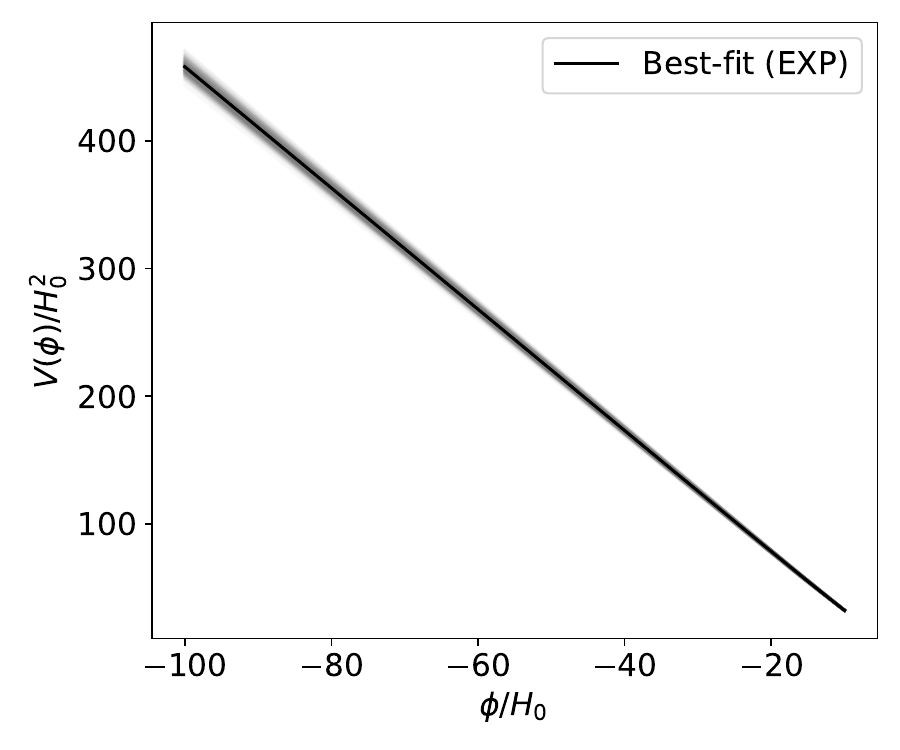}
\hspace{0.01in}
\includegraphics[scale=0.35]{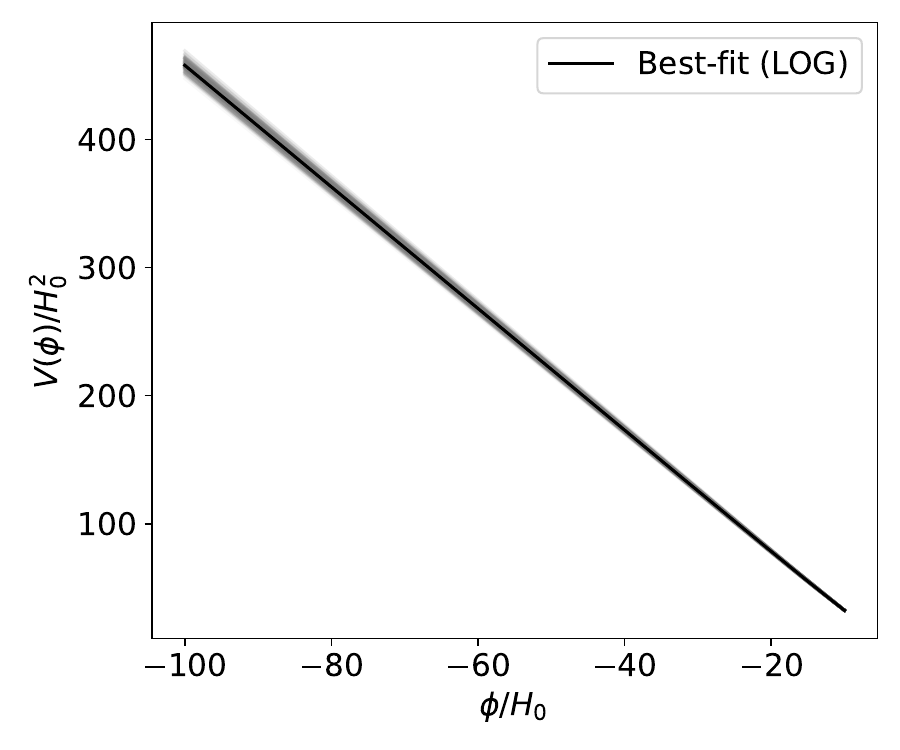}
\hspace{0.01in}
\includegraphics[scale=0.35]{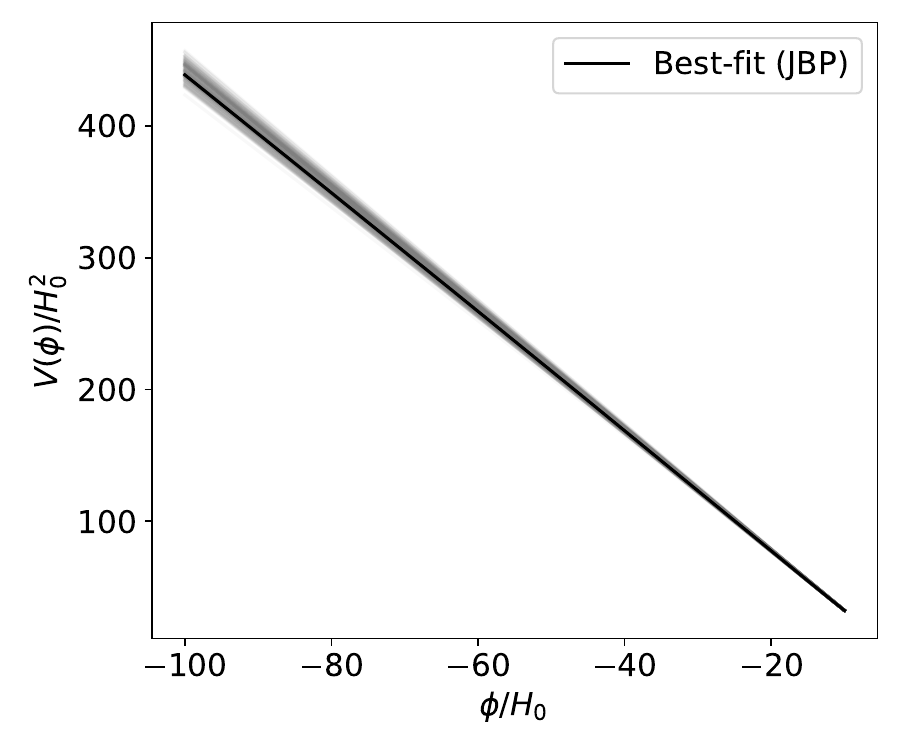}
\hspace{0.01in}
\includegraphics[scale=0.35]{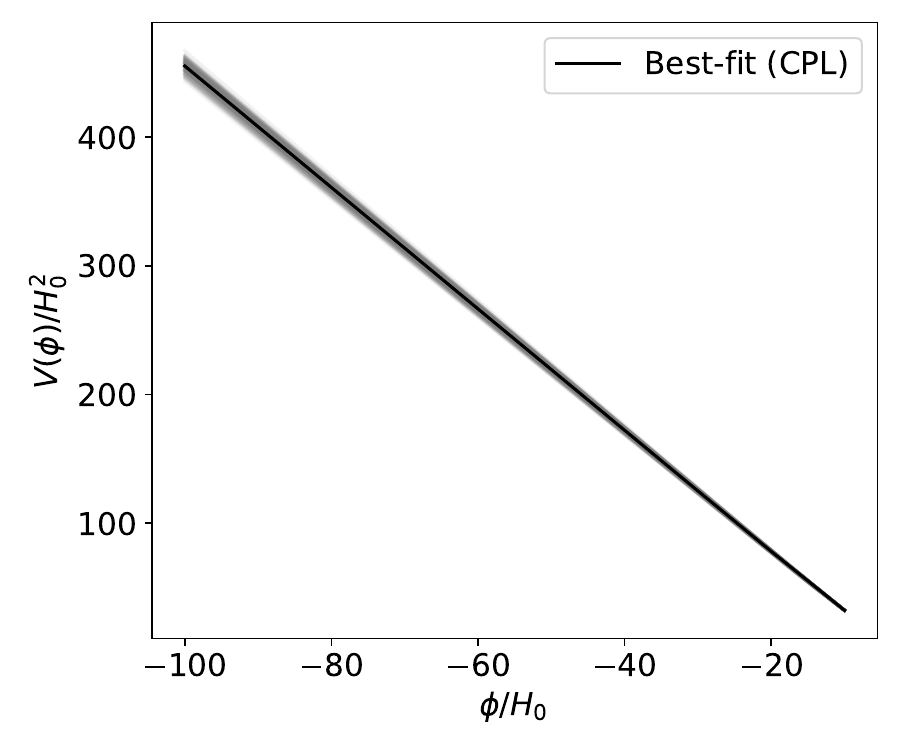}
\caption{The reconstructed potential $V$ normalized to $H_0^2$ for all the cases as a function of $\phi$ normalized to $H_0$. The black line corresponds to the best fit. The grey lines have been randomly sampled from the chains, and for each of them, we have solved Eq.\ \eqref{eq:z_to_phi} to map $z$ to $\phi$.}
\label{fig:v}

\end{figure*}

\begin{figure*}

\includegraphics[scale=0.4]{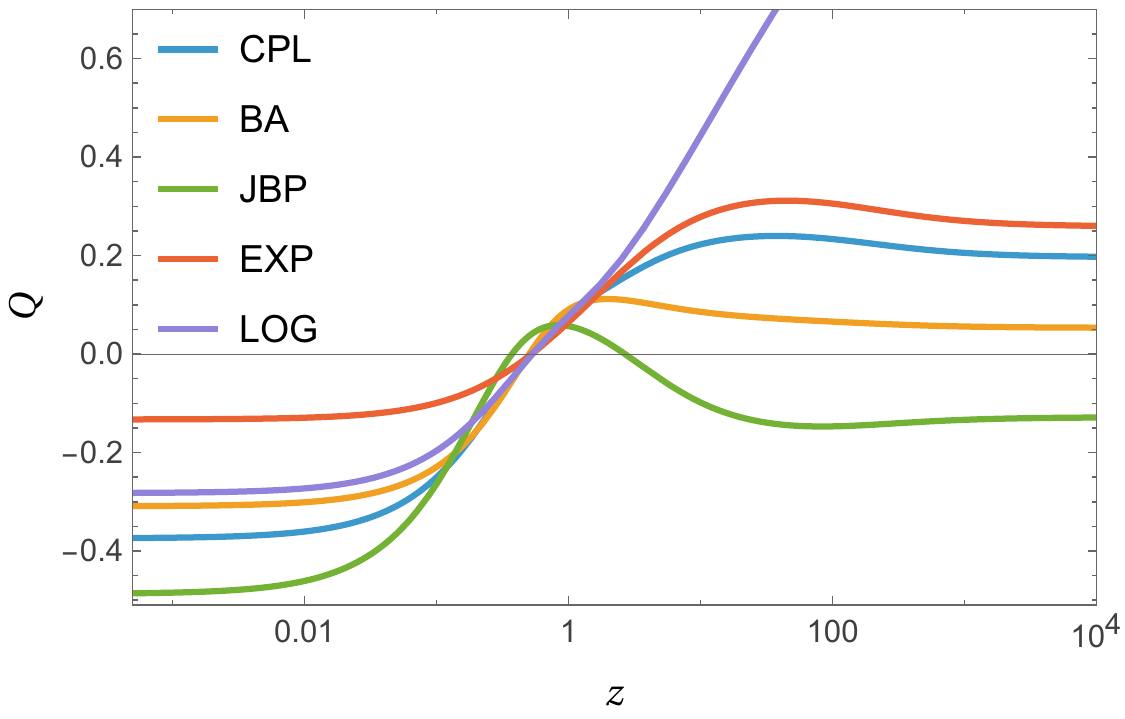} 
\hspace{0.01in}
\caption{The function $Q$ shows the difference of each model from $\Lambda$CDM (for which $Q=0$). The phantom behavior ($w_{\rm de} < -1$) corresponds to $Q$ becoming positive. For the LOG model, $Q$ is not bounded, since $w_{\rm de}$, for this model, is also not bounded. Furthermore, for the best-fit JBP model, the model becomes phantom only in a finite range of redshifts.}
\label{fig:Q}

\end{figure*}

Here, we reconstruct the potential $V(\phi)$ as a function of the redshift $z$ and present some concrete examples. To begin, we rewrite Eq.~\eqref{eq:dotphi1} as
\begin{equation}
    \frac{{\rm d}z}{{\rm d}\phi}=-(1+z)\,H\,\frac{2\Mpl^2}{3\,(\rho+P)}\,.\label{eq:z_to_phi}
\end{equation}
From the perspective of the MCMC analysis, it is convenient to express the dark energy density, $\rho_{\rm de}$ explicitly as a given function of redshift, such that $\rho_{\rm de}=\rho_{\rm de}(z)$. Using Eq.\ \eqref{VCDM1}, the Hubble function $H$ can be regarded as a known function of $z$, since the continuity equations \eqref{eq:cont_mat} for each standard matter fields are employed to determine $\rho_i(z)$ (and $P_i(z)$). Consequently, the right-hand side of \eqref{eq:z_to_phi} becomes a known function of $z$. We can now integrate \eqref{eq:z_to_phi}, setting an arbitrary initial condition, for instance, $\phi(z=0)=-10\,H(z=0)$. Solving this differential equation yields $z=z(\phi)$, which can subsequently be substituted into Eq.\ \eqref{eq:Einst1} to reconstruct the potential $V(\phi)$.
\begin{equation}
 V(\phi)=\frac13\,\phi^2-\frac{\rho\bigl(z(\phi)\bigr)}{\Mpl^2}\,.
\end{equation}
For all the cases considered, the reconstructed potentials closely resemble that of the BA model, as shown in Fig.\ \ref{fig:v}. To show the deviations for each model from $\Lambda$CDM, we introduce in Fig.\ \ref{fig:Q} the following dimensionless quantity
\begin{equation}
    Q\equiv\frac{V_{,\phi\phi}\, (\frac12V_{,\phi}-\frac{\phi}{3})^2}{V-\phi V_{,\phi}+\frac34\,V_{,\phi}^2}\,.
\end{equation}
In fact, $Q$ exactly vanishes in the $\Lambda$CDM limit, and has the nice property that it is built of the potential and its derivatives. It can be rewritten as $Q=V_{,\phi\phi} H^2/(\rho_{\rm de}/\Mpl^2)=V_{,\phi\phi}/(3\Omega_{\rm de})$. For finite but non-zero values of $Q$, we have $V_{,\phi\phi}\propto \Omega_{\rm de}$, which explains why the reconstructed potentials in Fig.\ \ref{fig:v} tend to look like straight lines (more and more, at large redshifts).

\bibliography{Ref}

@article{Planck:2018vyg,
    author = "Aghanim, N. and others",
    collaboration = "Planck",
    title = "{Planck 2018 results. VI. Cosmological parameters}",
    eprint = "1807.06209",
    archivePrefix = "arXiv",
    primaryClass = "astro-ph.CO",
    doi = "10.1051/0004-6361/201833910",
    journal = "Astron. Astrophys.",
    volume = "641",
    pages = "A6",
    year = "2020",
    note = "[Erratum: Astron.Astrophys. 652, C4 (2021)]"
}

@article{Riess:2019cxk,
    author = "Riess, Adam G. and Casertano, Stefano and Yuan, Wenlong and Macri, Lucas M. and Scolnic, Dan",
    title = "{Large Magellanic Cloud Cepheid Standards Provide a 1{\%} Foundation for the Determination of the Hubble Constant and Stronger Evidence for Physics beyond $\Lambda$CDM}",
    eprint = "1903.07603",
    archivePrefix = "arXiv",
    primaryClass = "astro-ph.CO",
    doi = "10.3847/1538-4357/ab1422",
    journal = "Astrophys. J.",
    volume = "876",
    number = "1",
    pages = "85",
    year = "2019"
}

@article{H0LiCOW:2019pvv,
    author = "Wong, Kenneth C. and others",
    collaboration = "H0LiCOW",
    title = "{H0LiCOW {\textendash} XIII. A 2.4 per cent measurement of H0 from lensed quasars: 5.3{\ensuremath{\sigma}} tension between early- and late-Universe probes}",
    eprint = "1907.04869",
    archivePrefix = "arXiv",
    primaryClass = "astro-ph.CO",
    doi = "10.1093/mnras/stz3094",
    journal = "Mon. Not. Roy. Astron. Soc.",
    volume = "498",
    number = "1",
    pages = "1420--1439",
    year = "2020"
}

@article{DiValentino:2021izs,
    author = "Di Valentino, Eleonora and Mena, Olga and Pan, Supriya and Visinelli, Luca and Yang, Weiqiang and Melchiorri, Alessandro and Mota, David F. and Riess, Adam G. and Silk, Joseph",
    title = "{In the realm of the Hubble tension{\textemdash}a review of solutions}",
    eprint = "2103.01183",
    archivePrefix = "arXiv",
    primaryClass = "astro-ph.CO",
    reportNumber = "IPPP/20/108",
    doi = "10.1088/1361-6382/ac086d",
    journal = "Class. Quant. Grav.",
    volume = "38",
    number = "15",
    pages = "153001",
    year = "2021"
}

@article{Perivolaropoulos:2021jda,
    author = "Perivolaropoulos, Leandros and Skara, Foteini",
    title = "{Challenges for {\ensuremath{\Lambda}}CDM: An update}",
    eprint = "2105.05208",
    archivePrefix = "arXiv",
    primaryClass = "astro-ph.CO",
    doi = "10.1016/j.newar.2022.101659",
    journal = "New Astron. Rev.",
    volume = "95",
    pages = "101659",
    year = "2022"
}

@article{Abdalla:2022yfr,
    author = "Abdalla, Elcio and others",
    title = "{Cosmology intertwined: A review of the particle physics, astrophysics, and cosmology associated with the cosmological tensions and anomalies}",
    eprint = "2203.06142",
    archivePrefix = "arXiv",
    primaryClass = "astro-ph.CO",
    reportNumber = "FERMILAB-CONF-22-192-SCD",
    doi = "10.1016/j.jheap.2022.04.002",
    journal = "JHEAp",
    volume = "34",
    pages = "49--211",
    year = "2022"
}

@article{Kamionkowski:2022pkx,
    author = "Kamionkowski, Marc and Riess, Adam G.",
    title = "{The Hubble Tension and Early Dark Energy}",
    eprint = "2211.04492",
    archivePrefix = "arXiv",
    primaryClass = "astro-ph.CO",
    doi = "10.1146/annurev-nucl-111422-024107",
    journal = "Ann. Rev. Nucl. Part. Sci.",
    volume = "73",
    pages = "153--180",
    year = "2023"
}

@article{DiValentino:2020zio,
    author = "Di Valentino, Eleonora and others",
    title = "{Snowmass2021 - Letter of interest cosmology intertwined II: The hubble constant tension}",
    eprint = "2008.11284",
    archivePrefix = "arXiv",
    primaryClass = "astro-ph.CO",
    reportNumber = "FERMILAB-PUB-21-590-PPD",
    doi = "10.1016/j.astropartphys.2021.102605",
    journal = "Astropart. Phys.",
    volume = "131",
    pages = "102605",
    year = "2021"
}

@article{Wu:2021jyk,
    author = "Wu, Qin and Zhang, Guo-Qiang and Wang, Fa-Yin",
    title = "{An 8~per{\,}cent determination of the Hubble constant from localized fast radio bursts}",
    eprint = "2108.00581",
    archivePrefix = "arXiv",
    primaryClass = "astro-ph.CO",
    doi = "10.1093/mnrasl/slac022",
    journal = "Mon. Not. Roy. Astron. Soc.",
    volume = "515",
    number = "1",
    pages = "L1--L5",
    year = "2022",
    note = "[Erratum: Mon.Not.Roy.Astron.Soc. 531, L8 (2024)]"
}

@article{Scolnic:2023mrv,
    author = "Scolnic, D. and Riess, A. G. and Wu, J. and Li, S. and Anand, G. S. and Beaton, R. and Casertano, S. and Anderson, R. I. and Dhawan, S. and Ke, X.",
    title = "{CATS: The Hubble Constant from Standardized TRGB and Type Ia Supernova Measurements}",
    eprint = "2304.06693",
    archivePrefix = "arXiv",
    primaryClass = "astro-ph.CO",
    doi = "10.3847/2041-8213/ace978",
    journal = "Astrophys. J. Lett.",
    volume = "954",
    number = "1",
    pages = "L31",
    year = "2023"
}

@article{Riess:2024vfa,
    author = "Riess, Adam G. and others",
    title = "{JWST Validates HST Distance Measurements: Selection of Supernova Subsample Explains Differences in JWST Estimates of Local H $_{0}$}",
    eprint = "2408.11770",
    archivePrefix = "arXiv",
    primaryClass = "astro-ph.CO",
    doi = "10.3847/1538-4357/ad8c21",
    journal = "Astrophys. J.",
    volume = "977",
    number = "1",
    pages = "120",
    year = "2024"
}

@article{Scolnic:2024hbh,
    author = "Scolnic, Daniel and others",
    title = "{The Hubble Tension in Our Own Backyard: DESI and the Nearness of the Coma Cluster}",
    eprint = "2409.14546",
    archivePrefix = "arXiv",
    primaryClass = "astro-ph.CO",
    doi = "10.3847/2041-8213/ada0bd",
    journal = "Astrophys. J. Lett.",
    volume = "979",
    number = "1",
    pages = "L9",
    year = "2025"
}

@article{Verde:2019ivm,
    author = "Verde, L. and Treu, T. and Riess, A. G.",
    title = "{Tensions between the Early and the Late Universe}",
    eprint = "1907.10625",
    archivePrefix = "arXiv",
    primaryClass = "astro-ph.CO",
    doi = "10.1038/s41550-019-0902-0",
    journal = "Nature Astron.",
    volume = "3",
    pages = "891",
    year = "2019"
}

@article{Linder:2002et,
    author = "Linder, Eric V.",
    title = "{Exploring the expansion history of the universe}",
    eprint = "astro-ph/0208512",
    archivePrefix = "arXiv",
    doi = "10.1103/PhysRevLett.90.091301",
    journal = "Phys. Rev. Lett.",
    volume = "90",
    pages = "091301",
    year = "2003"
}

@article{Jassal:2004ej,
    author = "Jassal, H. K. and Bagla, J. S. and Padmanabhan, T.",
    title = "{WMAP constraints on low redshift evolution of dark energy}",
    eprint = "astro-ph/0404378",
    archivePrefix = "arXiv",
    doi = "10.1111/j.1745-3933.2005.08577.x",
    journal = "Mon. Not. Roy. Astron. Soc.",
    volume = "356",
    pages = "L11--L16",
    year = "2005"
}

@article{Alam:2004jy,
    author = "Alam, Ujjaini and Sahni, Varun and Starobinsky, A. A.",
    title = "{The Case for dynamical dark energy revisited}",
    eprint = "astro-ph/0403687",
    archivePrefix = "arXiv",
    doi = "10.1088/1475-7516/2004/06/008",
    journal = "JCAP",
    volume = "06",
    pages = "008",
    year = "2004"
}

@article{Pan:2017zoh,
    author = "Pan, Supriya and Saridakis, Emmanuel N. and Yang, Weiqiang",
    title = "{Observational Constraints on Oscillating Dark-Energy Parametrizations}",
    eprint = "1712.05746",
    archivePrefix = "arXiv",
    primaryClass = "astro-ph.CO",
    reportNumber = "Phys.Rev. D98 (2018) no.6, 063510",
    doi = "10.1103/PhysRevD.98.063510",
    journal = "Phys. Rev. D",
    volume = "98",
    number = "6",
    pages = "063510",
    year = "2018"
}

@article{DeFelice:2012vd,
    author = "De Felice, Antonio and Nesseris, Savvas and Tsujikawa, Shinji",
    title = "{Observational constraints on dark energy with a fast varying equation of state}",
    eprint = "1203.6760",
    archivePrefix = "arXiv",
    primaryClass = "astro-ph.CO",
    doi = "10.1088/1475-7516/2012/05/029",
    journal = "JCAP",
    volume = "05",
    pages = "029",
    year = "2012"
}

@article{DiValentino:2019dzu,
    author = "Di Valentino, Eleonora and Melchiorri, Alessandro and Silk, Joseph",
    title = "{Cosmological constraints in extended parameter space from the Planck 2018 Legacy release}",
    eprint = "1908.01391",
    archivePrefix = "arXiv",
    primaryClass = "astro-ph.CO",
    doi = "10.1088/1475-7516/2020/01/013",
    journal = "JCAP",
    volume = "01",
    pages = "013",
    year = "2020"
}

@article{Li:2019yem,
    author = "Li, Xiaolei and Shafieloo, Arman",
    title = "{A Simple Phenomenological Emergent Dark Energy Model can Resolve the Hubble Tension}",
    eprint = "1906.08275",
    archivePrefix = "arXiv",
    primaryClass = "astro-ph.CO",
    doi = "10.3847/2041-8213/ab3e09",
    journal = "Astrophys. J. Lett.",
    volume = "883",
    number = "1",
    pages = "L3",
    year = "2019"
}

@article{Escamilla:2023oce,
    author = "Escamilla, Luis A. and Giar{\`e}, William and Di Valentino, Eleonora and Nunes, Rafael C. and Vagnozzi, Sunny",
    title = "{The state of the dark energy equation of state circa 2023}",
    eprint = "2307.14802",
    archivePrefix = "arXiv",
    primaryClass = "astro-ph.CO",
    doi = "10.1088/1475-7516/2024/05/091",
    journal = "JCAP",
    volume = "05",
    pages = "091",
    year = "2024"
}

@article{Giare:2024oil,
    author = "Giar{\`e}, William",
    title = "{Dynamical dark energy beyond Planck? Constraints from multiple CMB probes, DESI BAO, and type-Ia supernovae}",
    eprint = "2409.17074",
    archivePrefix = "arXiv",
    primaryClass = "astro-ph.CO",
    doi = "10.1103/ss37-cxhn",
    journal = "Phys. Rev. D",
    volume = "112",
    number = "2",
    pages = "023508",
    year = "2025"
}

@article{DESI:2024mwx,
    author = "Adame, A. G. and others",
    collaboration = "DESI",
    title = "{DESI 2024 VI: cosmological constraints from the measurements of baryon acoustic oscillations}",
    eprint = "2404.03002",
    archivePrefix = "arXiv",
    primaryClass = "astro-ph.CO",
    reportNumber = "FERMILAB-PUB-24-0154-PPD",
    doi = "10.1088/1475-7516/2025/02/021",
    journal = "JCAP",
    volume = "02",
    pages = "021",
    year = "2025"
}

@article{Cortes:2024lgw,
    author = "Cort{\^e}s, Marina and Liddle, Andrew R.",
    title = "{Interpreting DESI's evidence for evolving dark energy}",
    eprint = "2404.08056",
    archivePrefix = "arXiv",
    primaryClass = "astro-ph.CO",
    doi = "10.1088/1475-7516/2024/12/007",
    journal = "JCAP",
    volume = "12",
    pages = "007",
    year = "2024"
}

@article{Park:2024vrw,
    author = "Park, Chan-Gyung and de Cruz P{\'e}rez, Javier and Ratra, Bharat",
    title = "{Using non-DESI data to confirm and strengthen the DESI 2024 spatially flat w0waCDM cosmological parametrization result}",
    eprint = "2405.00502",
    archivePrefix = "arXiv",
    primaryClass = "astro-ph.CO",
    doi = "10.1103/PhysRevD.110.123533",
    journal = "Phys. Rev. D",
    volume = "110",
    number = "12",
    pages = "123533",
    year = "2024"
}

@article{Zheng:2024qzi,
    author = "Zheng, Jie and Qiang, Da-Chun and You, Zhi-Qiang",
    title = "{Cosmological constraints on dark energy models using DESI BAO 2024}",
    eprint = "2412.04830",
    archivePrefix = "arXiv",
    primaryClass = "astro-ph.CO",
    month = "12",
    year = "2024",
doi = "",
    journal = "",
}

@article{Nesseris:2025lke,
    author = "Nesseris, Savvas and Akrami, Yashar and Starkman, Glenn D.",
    title = "{To CPL, or not to CPL? What we have not learned about the dark energy equation of state}",
    eprint = "2503.22529",
    archivePrefix = "arXiv",
    primaryClass = "astro-ph.CO",
    reportNumber = "IFT-UAM/CSIC-25-29",
    month = "3",
    year = "2025",
doi = "",
    journal = "",
}

@article{DESI:2025zgx,
    author = "Abdul Karim, M. and others",
    collaboration = "DESI",
    title = "{DESI DR2 Results II: Measurements of Baryon Acoustic Oscillations and Cosmological Constraints}",
    eprint = "2503.14738",
    archivePrefix = "arXiv",
    primaryClass = "astro-ph.CO",
    reportNumber = "FERMILAB-PUB-25-0169-PPD",
    month = "3",
    year = "2025",
doi="",
journal = ""
}

@article{Bhattacharjee:2025xeb,
    author = "Bhattacharjee, Sauvik and Halder, Sudip and de Haro, Jaume and Pan, Supriya and Saridakis, Emmanuel N.",
    title = "{Accelerating Universe without dark energy: matter creation after DESI DR2}",
    eprint = "2507.15575",
    archivePrefix = "arXiv",
    primaryClass = "astro-ph.CO",
    month = "7",
    year = "2025",
doi="",
journal = ""
}

@article{Plaza:2025gcv,
    author = "Plaza, Francisco and Kraiselburd, Lucila",
    title = "{Testing f(R)-gravity models with DESI DR2 2025-BAO and other cosmological data}",
    doi = "10.1103/gtrg-56fj",
    journal = "Phys. Rev. D",
    volume = "112",
    number = "2",
    pages = "023554",
    year = "2025",
doi="",
journal = ""
}

@article{Mishra:2025goj,
    author = "Mishra, Swagat S. and Matthewson, William L. and Sahni, Varun and Shafieloo, Arman and Shtanov, Yuri",
    title = "{Braneworld Dark Energy in light of DESI DR2}",
    eprint = "2507.07193",
    archivePrefix = "arXiv",
    primaryClass = "astro-ph.CO",
    month = "7",
    year = "2025",
doi="",
journal = ""
}

@article{Du:2025xes,
    author = "Du, Guo-Hong and Li, Tian-Nuo and Wu, Peng-Ju and Zhang, Jing-Fei and Zhang, Xin",
    title = "{Cosmological Preference for a Positive Neutrino Mass at 2.7$\sigma$: A Joint Analysis of DESI DR2, DESY5, and DESY1 Data}",
    eprint = "2507.16589",
    archivePrefix = "arXiv",
    primaryClass = "astro-ph.CO",
    month = "7",
    year = "2025",
doi="",
journal = ""
}

@article{Li:2025ops,
    author = "Li, Jun-Xian and Wang, Shuang",
    title = "{Reconstructing dark energy with model independent methods after DESI DR2 BAO}",
    eprint = "2506.22953",
    archivePrefix = "arXiv",
    primaryClass = "astro-ph.CO",
    month = "6",
    year = "2025",
doi="",
journal = ""
}

@article{Lee:2025kbn,
    author = "Lee, Seokcheon",
    title = "{The Impact of $\Omega_{m0}$ Prior Bias on Cosmological Parameter Estimation: Reconciling DESI DR2 BAO and Pantheon+ SNe Data Combination Results}",
    eprint = "2506.16022",
    archivePrefix = "arXiv",
    primaryClass = "astro-ph.CO",
    month = "6",
    year = "2025",
doi="",
journal = ""
}

@article{Scherer:2025esj,
    author = "Scherer, Mateus and Sabogal, Miguel A. and Nunes, Rafael C. and De Felice, Antonio",
    title = "{Challenging $\Lambda$CDM: 5$\sigma$ Evidence for a Dynamical Dark Energy Late-Time Transition}",
    eprint = "2504.20664",
    archivePrefix = "arXiv",
    primaryClass = "astro-ph.CO",
    month = "4",
    year = "2025",
doi="",
journal = ""
}

@article{DeFelice:2020eju,
    author = "De Felice, Antonio and Doll, Andreas and Mukohyama, Shinji",
    title = "{A theory of type-II minimally modified gravity}",
    eprint = "2004.12549",
    archivePrefix = "arXiv",
    primaryClass = "gr-qc",
    reportNumber = "YITP-20-55, IPMU20-0040",
    doi = "10.1088/1475-7516/2020/09/034",
    journal = "JCAP",
    volume = "09",
    pages = "034",
    year = "2020"
}

@article{Ganz:2022zgs,
    author = "Ganz, Alexander and Martens, Paul and Mukohyama, Shinji and Namba, Ryo",
    title = "{Bouncing cosmology in VCDM}",
    eprint = "2212.13561",
    archivePrefix = "arXiv",
    primaryClass = "gr-qc",
    reportNumber = "RIKEN-iTHEMS-Report-22",
    doi = "10.1088/1475-7516/2023/04/060",
    journal = "JCAP",
    volume = "04",
    pages = "060",
    year = "2023"
}

@article{Ganz:2024ihb,
    author = "Ganz, Alexander and Martens, Paul and Mukohyama, Shinji and Namba, Ryo",
    title = "{Bispectrum from inflation/bouncing Universe in VCDM}",
    eprint = "2407.02882",
    archivePrefix = "arXiv",
    primaryClass = "gr-qc",
    reportNumber = "RIKEN-iTHEMS-Report-24",
    month = "7",
    year = "2024",
doi = "",
journal = ""
}

@article{DeFelice:2020cpt,
    author = "De Felice, Antonio and Mukohyama, Shinji and Pookkillath, Masroor C.",
    title = "{Addressing $H_0$ tension by means of VCDM}",
    eprint = "2009.08718",
    archivePrefix = "arXiv",
    primaryClass = "astro-ph.CO",
    reportNumber = "YITP-20-117, IPMU 20-0098",
    doi = "10.1016/j.physletb.2021.136201",
    journal = "Phys. Lett. B",
    volume = "816",
    pages = "136201",
    year = "2021",
    note = "[Erratum: Phys.Lett.B 818, 136364 (2021)]"
}

@article{Jalali:2023wqh,
    author = "Jalali, Atabak Fathe and Martens, Paul and Mukohyama, Shinji",
    title = "{Spherical scalar collapse in a type-II minimally modified gravity}",
    eprint = "2306.10672",
    archivePrefix = "arXiv",
    primaryClass = "gr-qc",
    reportNumber = "YITP-23-75, IPMU23-0023",
    doi = "10.1103/PhysRevD.109.044053",
    journal = "Phys. Rev. D",
    volume = "109",
    number = "4",
    pages = "044053",
    year = "2024"
}

@article{DeFelice:2022riv,
    author = "De Felice, Antonio and Maeda, Kei-ichi and Mukohyama, Shinji and Pookkillath, Masroor C.",
    title = "{Gravitational collapse and formation of a black hole in a type II minimally modified gravity theory}",
    eprint = "2211.14760",
    archivePrefix = "arXiv",
    primaryClass = "gr-qc",
    reportNumber = "YITP-22-142, IPMU22-0062",
    doi = "10.1088/1475-7516/2023/03/030",
    journal = "JCAP",
    volume = "03",
    pages = "030",
    year = "2023"
}

@article{DeFelice:2022uxv,
    author = "De Felice, Antonio and Maeda, Kei-ichi and Mukohyama, Shinji and Pookkillath, Masroor C.",
    title = "{Comparison of two theories of Type-IIa minimally modified gravity}",
    eprint = "2204.08294",
    archivePrefix = "arXiv",
    primaryClass = "gr-qc",
    reportNumber = "YITP-22-39, IPMU22-0020",
    doi = "10.1103/PhysRevD.106.024028",
    journal = "Phys. Rev. D",
    volume = "106",
    number = "2",
    pages = "024028",
    year = "2022"
}

@article{Chevallier:2000qy,
    author = "Chevallier, Michel and Polarski, David",
    title = "{Accelerating universes with scaling dark matter}",
    eprint = "gr-qc/0009008",
    archivePrefix = "arXiv",
    doi = "10.1142/S0218271801000822",
    journal = "Int. J. Mod. Phys. D",
    volume = "10",
    pages = "213--224",
    year = "2001"
}

@article{Wang:2025znm,
    author = "Wang, Jia-Qi and Cai, Rong-Gen and Guo, Zong-Kuan and Wang, Shao-Jiang",
    title = "{Resolving the Planck-DESI tension by non-minimally coupled quintessence}",
    eprint = "2508.01759",
    archivePrefix = "arXiv",
    primaryClass = "astro-ph.CO",
    month = "8",
    year = "2025",
doi = "",
journal = ""
}

@article{Li:2025dwz,
    author = "Li, Tian-Nuo and Wu, Peng-Ju and Du, Guo-Hong and Yao, Yan-Hong and Zhang, Jing-Fei and Zhang, Xin",
    title = "{Exploring non-cold dark matter in a scenario of dynamical dark energy with DESI DR2 data}",
    eprint = "2507.07798",
    archivePrefix = "arXiv",
    primaryClass = "astro-ph.CO",
    month = "7",
    year = "2025",
doi = "",
journal = ""
}

@article{Wu:2024faw,
    author = "Wu, Peng-Ju and Zhang, Xin",
    title = "{Measuring cosmic curvature with non-CMB observations}",
    eprint = "2411.06356",
    archivePrefix = "arXiv",
    primaryClass = "astro-ph.CO",
    month = "11",
    year = "2024",
doi = "",
journal = ""
}

@article{Wang:2025bkk,
    author = "Wang, Deng and Mota, David",
    title = "{Did DESI DR2 truly reveal dynamical dark energy?}",
    eprint = "2504.15222",
    archivePrefix = "arXiv",
    primaryClass = "astro-ph.CO",
    month = "4",
    year = "2025",
doi = "",
journal = ""
}

@article{Poulin:2025nfb,
    author = "Poulin, Vivian and Smith, Tristan L. and Calder{\'o}n, Rodrigo and Simon, Th{\'e}o",
    title = "{Impact of ACT DR6 and DESI DR2 for Early Dark Energy and the Hubble tension}",
    eprint = "2505.08051",
    archivePrefix = "arXiv",
    primaryClass = "astro-ph.CO",
    month = "5",
    year = "2025",
doi = "",
journal = ""
}

@article{Ling:2025lmw,
    author = "Ling, Jia-Le and Du, Guo-Hong and Li, Tian-Nuo and Zhang, Jing-Fei and Wang, Shao-Jiang and Zhang, Xin",
    title = "{Model-independent cosmological inference after the DESI DR2 data with improved inverse distance ladder}",
    eprint = "2505.22369",
    archivePrefix = "arXiv",
    primaryClass = "astro-ph.CO",
    month = "5",
    year = "2025",
doi = "",
journal = ""
}

@article{Lee:2022cyh,
    author = "Lee, Bum-Hoon and Lee, Wonwoo and Colg{\'a}in, Eoin {\'O}. and Sheikh-Jabbari, M. M. and Thakur, Somyadip",
    title = "{Is local H $_{0}$ at odds with dark energy EFT?}",
    eprint = "2202.03906",
    archivePrefix = "arXiv",
    primaryClass = "astro-ph.CO",
    doi = "10.1088/1475-7516/2022/04/004",
    journal = "JCAP",
    volume = "04",
    number = "04",
    pages = "004",
    year = "2022"
}

@article{Barboza:2008rh,
    author = "Barboza, Jr., E. M. and Alcaniz, J. S.",
    title = "{A parametric model for dark energy}",
    eprint = "0805.1713",
    archivePrefix = "arXiv",
    primaryClass = "astro-ph",
    doi = "10.1016/j.physletb.2008.08.012",
    journal = "Phys. Lett. B",
    volume = "666",
    pages = "415--419",
    year = "2008"
}

@article{Mehrabi:2018dru,
    author = "Mehrabi, A.",
    title = "{Growth of perturbations in dark energy parametrization scenarios}",
    eprint = "1804.09886",
    archivePrefix = "arXiv",
    primaryClass = "astro-ph.CO",
    doi = "10.1103/PhysRevD.97.083522",
    journal = "Phys. Rev. D",
    volume = "97",
    number = "8",
    pages = "083522",
    year = "2018"
}

@article{Jassal:2005qc,
    author = "Jassal, Harvinder Kaur and Bagla, J. S. and Padmanabhan, T.",
    title = "{Observational constraints on low redshift evolution of dark energy: How consistent are different observations?}",
    eprint = "astro-ph/0506748",
    archivePrefix = "arXiv",
    doi = "10.1103/PhysRevD.72.103503",
    journal = "Phys. Rev. D",
    volume = "72",
    pages = "103503",
    year = "2005"
}

@article{Pan:2019brc,
    author = "Pan, Supriya and Yang, Weiqiang and Paliathanasis, Andronikos",
    title = "{Imprints of an extended Chevallier{\textendash}Polarski{\textendash}Linder parametrization on the large scale of our universe}",
    eprint = "1902.07108",
    archivePrefix = "arXiv",
    primaryClass = "astro-ph.CO",
    doi = "10.1140/epjc/s10052-020-7832-y",
    journal = "Eur. Phys. J. C",
    volume = "80",
    number = "3",
    pages = "274",
    year = "2020"
}

@article{Najafi:2024qzm,
    author = "Najafi, Mahdi and Pan, Supriya and Di Valentino, Eleonora and Firouzjaee, Javad T.",
    title = "{Dynamical dark energy confronted with multiple CMB missions}",
    eprint = "2407.14939",
    archivePrefix = "arXiv",
    primaryClass = "astro-ph.CO",
    doi = "10.1016/j.dark.2024.101539",
    journal = "Phys. Dark Univ.",
    volume = "45",
    pages = "101539",
    year = "2024"
}

@article{Tripathi:2016slv,
    author = "Tripathi, Ashutosh and Sangwan, Archana and Jassal, H. K.",
    title = "{Dark energy equation of state parameter and its evolution at low redshift}",
    eprint = "1611.01899",
    archivePrefix = "arXiv",
    primaryClass = "astro-ph.CO",
    doi = "10.1088/1475-7516/2017/06/012",
    journal = "JCAP",
    volume = "06",
    pages = "012",
    year = "2017"
}

@article{Efstathiou:1999tm,
    author = "Efstathiou, G.",
    title = "{Constraining the equation of state of the universe from distant type Ia supernovae and cosmic microwave background anisotropies}",
    eprint = "astro-ph/9904356",
    archivePrefix = "arXiv",
    doi = "10.1046/j.1365-8711.1999.02997.x",
    journal = "Mon. Not. Roy. Astron. Soc.",
    volume = "310",
    pages = "842--850",
    year = "1999"
}

@article{Yang:2021flj,
    author = "Yang, Weiqiang and Di Valentino, Eleonora and Pan, Supriya and Wu, Yabo and Lu, Jianbo",
    title = "{Dynamical dark energy after Planck CMB final release and $H_0$ tension}",
    eprint = "2101.02168",
    archivePrefix = "arXiv",
    primaryClass = "astro-ph.CO",
    doi = "10.1093/mnras/staa3914",
    journal = "Mon. Not. Roy. Astron. Soc.",
    volume = "501",
    number = "4",
    pages = "5845--5858",
    year = "2021"
}

@article{Planck:2019nip,
    author = "Aghanim, N. and others",
    collaboration = "Planck",
    title = "{Planck 2018 results. V. CMB power spectra and likelihoods}",
    eprint = "1907.12875",
    archivePrefix = "arXiv",
    primaryClass = "astro-ph.CO",
    doi = "10.1051/0004-6361/201936386",
    journal = "Astron. Astrophys.",
    volume = "641",
    pages = "A5",
    year = "2020"
}

@article{DESI:2025fii,
    author = "Lodha, K. and others",
    collaboration = "DESI",
    title = "{Extended Dark Energy analysis using DESI DR2 BAO measurements}",
    eprint = "2503.14743",
    archivePrefix = "arXiv",
    primaryClass = "astro-ph.CO",
    reportNumber = "FERMILAB-PUB-25-0164-PPD",
    month = "3",
    year = "2025",
      doi  = " ",
      journal =" "
}

@article{DESI:2025qqy,
    author = "Andrade, U. and others",
    collaboration = "DESI",
    title = "{Validation of the DESI DR2 Measurements of Baryon Acoustic Oscillations from Galaxies and Quasars}",
    eprint = "2503.14742",
    archivePrefix = "arXiv",
    primaryClass = "astro-ph.CO",
    reportNumber = "FERMILAB-PUB-25-0162-PPD",
    month = "3",
    year = "2025",
doi ="",
journal = ""
}

@article{Blas:2011rf,
    author = "Blas, Diego and Lesgourgues, Julien and Tram, Thomas",
    title = "{The Cosmic Linear Anisotropy Solving System (CLASS) II: Approximation schemes}",
    eprint = "1104.2933",
    archivePrefix = "arXiv",
    primaryClass = "astro-ph.CO",
    reportNumber = "CERN-PH-TH-2011-082, LAPTH-010-11",
    doi = "10.1088/1475-7516/2011/07/034",
    journal = "JCAP",
    volume = "07",
    pages = "034",
    year = "2011"
}

@article{Brinckmann:2018cvx,
    author = "Brinckmann, Thejs and Lesgourgues, Julien",
    title = "{MontePython 3: boosted MCMC sampler and other features}",
    eprint = "1804.07261",
    archivePrefix = "arXiv",
    primaryClass = "astro-ph.CO",
    reportNumber = "TTK-18-15",
    doi = "10.1016/j.dark.2018.100260",
    journal = "Phys. Dark Univ.",
    volume = "24",
    pages = "100260",
    year = "2019"
}

@article{Lewis:2019xzd,
    author = "Lewis, Antony",
    title = "{GetDist: a Python package for analysing Monte Carlo samples}",
    eprint = "1910.13970",
    archivePrefix = "arXiv",
    primaryClass = "astro-ph.IM",
    month = "10",
    year = "2019",
doi="",
journal = ""
}

@article{DeFelice:2020onz,
    author = "De Felice, Antonio and Doll, Andreas and Larrouturou, Fran{\c{c}}ois and Mukohyama, Shinji",
    title = "{Black holes in a type-II minimally modified gravity}",
    eprint = "2010.13067",
    archivePrefix = "arXiv",
    primaryClass = "gr-qc",
    reportNumber = "YITP-20-131, IPMU20-0111",
    doi = "10.1088/1475-7516/2021/03/004",
    journal = "JCAP",
    volume = "03",
    pages = "004",
    year = "2021"
}

@article{DeFelice:2021xps,
    author = "De Felice, Antonio and Mukohyama, Shinji and Pookkillath, Masroor C.",
    title = "{Static, spherically symmetric objects in type-II minimally modified gravity}",
    eprint = "2110.14496",
    archivePrefix = "arXiv",
    primaryClass = "gr-qc",
    reportNumber = "YITP-21-127, IPMU21-0068",
    doi = "10.1103/PhysRevD.105.104013",
    journal = "Phys. Rev. D",
    volume = "105",
    number = "10",
    pages = "104013",
    year = "2022"
}

@article{Akarsu:2024qsi,
    author = {Akarsu, {\"O}zg{\"u}r and De Felice, Antonio and Di Valentino, Eleonora and Kumar, Suresh and Nunes, Rafael C. and {\"O}z{\"u}lker, Emre and Vazquez, J. Alberto and Yadav, Anita},
    title = "{$\Lambda_{\rm s}$CDM cosmology from a type-II minimally modified gravity}",
    eprint = "2402.07716",
    archivePrefix = "arXiv",
    primaryClass = "astro-ph.CO",
    reportNumber = "YITP-24-18",
    month = "2",
    year = "2024",
doi = "",
journal = ""
}

\end{document}